\documentclass[reprint,amsmath,amssymb,aps,prl,superscriptaddress]{revtex4-2}
\usepackage{graphicx}
\usepackage{dcolumn}
\usepackage{bm}
\usepackage{color}
\usepackage{units}
\usepackage{wasysym}
\usepackage{MnSymbol}
\usepackage{comment}
\usepackage{times}
\usepackage{soul}
\usepackage{upgreek}
\usepackage{amssymb}
\usepackage[unicode=true,pdfusetitle,bookmarks=false,colorlinks=true,citecolor=blue,urlcolor=blue,linkcolor=blue]{hyperref}

\begin{document}

\title{Emergent nonreciprocity in open thermodynamically-consistent chemical reaction networks}
\author{Daniel Evans}
\affiliation{Department of Materials Science and Engineering, University of California, Berkeley, California 94720, USA}
\affiliation{Materials Sciences Division, Lawrence Berkeley National Laboratory, Berkeley, California 94720, USA}
\author{Yizhi Shen}
\affiliation{Applied Mathematics and Computational Research Division, Lawrence Berkeley National Laboratory, Berkeley, CA 94720, USA}
\author{Ahmad K. Omar}
\email{aomar@berkeley.edu}
\affiliation{Department of Materials Science and Engineering, University of California, Berkeley, California 94720, USA}
\affiliation{Materials Sciences Division, Lawrence Berkeley National Laboratory, Berkeley, California 94720, USA}

\begin{abstract}
Nonreciprocity, a hallmark of nonequilibrium systems, can generate dynamics not possible near thermodynamic equilibrium, including oscillatory and rotating patterns. 
The onset of temporal oscillations is often evident in linearized dynamics, where nonreciprocity appears as complex eigenvalues of an asymmetric Jacobian.
Here, we show that the topology of open, thermodynamically-consistent chemical reaction networks can result in oscillatory instabilities near nonequilibrium steady states. 
These instabilities arise from chemostat-induced breaking of Onsager reciprocity, while the local equilibrium hypothesis preserves the variational structure of the dissipative part of the dynamics. 
Numerical results confirm that such nonreciprocity in reaction-diffusion systems produces oscillatory dynamics that nevertheless minimize a free energy.
\end{abstract}
\maketitle

\textit{Introduction.--} Temporal oscillations in the concentrations of chemical species pervade living processes, from cytosolic calcium waves in eukaryotic cells~\cite{Berridge1988, Fewtrell1993, Xiong2025} to pole-to-pole protein cycling on the membrane of \textit{E. coli}~\cite{raskin1999rapid, raskin1999minde, hu1999topological, vecchiarelli2016membrane}. Many such behaviors are captured by dynamical descriptions of the relevant fields involving reactions and/or diffusion.
Temporal oscillations of these fields can arise, for example, from microscopic nonreciprocal forces (such as self-propulsion or predator-prey interactions between particles)~\cite{Ivlev2015, Agudo2019, Fruchart2021, Chiu2023, Duan2023, Duan2025, Kreienkamp2024, Kreienkamp20242, Navas2024, Chen2024, Stenhammer2015, Wysocki2016, Wittkowski2017, Agrawal2017, Banjeree2022, Dinelli2023, Tucci2024, Mason2024}. This has motivated the proposal of phenomenological continuum models, including the nonreciprocal Cahn-Hilliard model~\cite{Saha2020, You2020, Brauns2024}. While these recent efforts have lent important insights into temporal oscillations in driven systems, a complete understanding of the relationship between the form of nonequilibrium driving and the resulting oscillations remains outstanding.

The onset of temporal oscillations from a spatially uniform steady state can be predicted through a linear stability analysis of the relevant fields.
In a system composed of $n_e$ chemical species, the common fields of interest are the coarse-grained number density fields, which we can compactly collect into a vector $\boldsymbol{\rho}$ of length $n_e$.
Near a steady state, the linearized dynamics of the spatially Fourier-transformed density fluctuations, $\delta \tilde{\boldsymbol{\rho}}_q$, will take the form ~$\partial_t \delta \tilde{\boldsymbol{\rho}}_q = \mathbf{A}(q) \cdot \delta \tilde{\boldsymbol{\rho}}_q$ where the Jacobian, $\mathbf{A}(q)$, is an $n_e \times n_e$ real matrix that depends on the magnitude of the wavevector $\mathbf{q}$, $q \equiv |\mathbf{q}|$.
When the eigenvalues of $\mathbf{A}$ are complex, we can expect temporal oscillations of the density fields that are typically associated with nonreciprocity. 
Complex eigenvalues require an asymmetric Jacobian, and it is perhaps for this reason that asymmetry is often considered to be a hallmark of nonreciprocity at the field level~\footnote{When considering the dynamics of fields which can take complex values, non-reciprocity is often believed to correspond to a non-Hermitian Jacobian. This again is not a sufficient definition of nonreciprocity, for the same reasons outlined in the main text.}.
However, asymmetry alone is not sufficient: many asymmetric matrices admit purely real spectra.
Most notably, passive systems near states of thermodynamic equilibrium generally have asymmetric Jacobians but are constrained to have purely real eigenvalues due to the variational structure of the dynamics.
We emphasize that the linearized dynamics are only \textit{genuinely nonreciprocal} when $\mathbf{A}$ has complex eigenvalues~\cite{fruchart2026}.

The above discussion may appear to suggest that temporal oscillations in the species concentrations and a free energy-minimizing principle are mutually exclusive.
As we will show, this is not the case -- passive dynamics are a particular, restricted form of free energy-minimizing dynamics that also precludes temporal oscillations.
To demonstrate this, we consider open, thermodynamically-consistent reaction-diffusion systems.
Even upon assuming local equilibrium, the topology of the reaction network and use of \textit{autonomous} chemostats (i.e.,~chemostats that instantaneously equilibrate the species chemical potential to that of a reservoir) on a subset of species can drive these systems to nonequilibrium steady states.
We find that a subclass of reaction networks -- those that admit \textit{complex-balanced}~\cite{Horn1972, Feinberg1972, Avanzini2021, Avanzini2024} steady states -- exhibit oscillatory instabilities (i.e.,~$\mathbf{A}$ has complex eigenvalues) while retaining the grand potential as a Lyapunov functional~\cite{Avanzini2024}.
In other words, the dynamics of these systems are genuinely nonreciprocal while acting to reduce a free energy.
We anticipate that the dynamics examined in this work belong to a broader class of driven systems that can robustly display temporal oscillations in the presence of a Lyapunov functional.

\textit{Constitutive Dynamics.--} We consider an open system of $n_e$ dynamically evolving species, whose concentrations are described by the vector $\boldsymbol{\rho}$, that undergo $n_r$ reactions in a fixed volume $V$ at temperature $T$.
While these species can be transformed through chemical reactions, they cannot be exchanged with the surrounding environment. 
We consider an additional $n_c$ species that are subject to chemostats~\cite{Remlein2025}. 
The system can exchange these chemostatted species with a reservoir of fixed chemical potential $\mu_i^{\rm chemo}$ for the $i$th species.
We treat these species as ideal (non-interacting) and ``autonomous'' (i.e.,~instantaneously~\cite{Remlein2025} relaxing to $\mu_i^{\rm chemo}$) for simplicity.
In the Supplemental Material (SM)~\footnote{see the Supplemental Material (SM) for supporting examples and derivations, which includes discussions related to Refs.~\cite{murugan2017topologically, Tang2021, Ezawa2022, Nelson2024, zheng2024topological, veenstra2024non, Belyansky2025}}, following Refs.~\cite{Polettini2014, Rao2016}, we distinguish between scenarios in which chemostatting results in equilibrium steady states and scenarios where it drives the system out of equilibrium.
The reactions can be expressed as:
\begin{equation}
    \boldsymbol{\nu}^+ \cdot \mathbf{Z} + \boldsymbol{\nu}^{+, \rm chemo} \cdot \mathbf{Z}^{\rm chemo}  \rightleftharpoons \boldsymbol{\nu}^- \cdot \mathbf{Z} + \boldsymbol{\nu}^{-, \rm chemo} \cdot \mathbf{Z}^{\rm chemo} ,
\end{equation}
where $\mathbf{Z}$ and $\mathbf{Z}^{\rm chemo}$ are vectors of length $n_e$ and $n_c$, respectively, holding the symbol of each evolving and chemostatted species.
We use a simple notation where addition between two symbols, $A$ and $B$, yields the string of characters $A+B$.
Additionally, $\boldsymbol{\nu}^{\pm}$ and $\boldsymbol{\nu}^{\pm, \rm chemo}$ are $n_r \times n_e$ and $n_r \times n_c$ matrices that encode the stoichiometry of each evolving and chemostatted species in the $\pm$ direction of reaction $r$.
These reactions enter the dynamics of $\boldsymbol{\rho}$ as generation terms in the species balance laws:
\begin{subequations}
\label{eq:gendyn}
\begin{equation}
    \partial_t \boldsymbol{\rho} = - \sum_{\alpha=1}^d \partial_\alpha \mathbf{J}^\alpha + \mathbf{j} \cdot \boldsymbol{\nu}, \label{eq:gendyn1}
\end{equation}
where $d$ is the number of spatial dimensions and $\partial_\alpha \equiv \partial / \partial \alpha$ is the spatial gradient in the $\alpha$th direction. 
The vector $\mathbf{J}^\alpha$ of length $n_e$ contains the fluxes
of all evolving species along the $\alpha$th direction (relative to the local velocity of a non-reacting and relatively immobile species that is not included in $n_e$). 
We have defined the net stoichiometry matrix as $\boldsymbol{\nu} \equiv \boldsymbol{\nu}^- - \boldsymbol{\nu}^+$. 
Finally, $\mathbf{j} \equiv \mathbf{j}^+ - \mathbf{j}^-$ is a length $n_r$ vector of net reaction rate densities which can be expressed as the difference in rate densities between the forward ($\mathbf{j}^+$) and backward ($\mathbf{j}^-$) reactions.

We now require constitutive equations for the species flux and reaction rates. 
We turn to the local equilibrium hypothesis which postulates that thermodynamic forces may still be used to describe nonequilibrium systems and is the basis of linear irreversible thermodynamics~\cite{degroot2013}.
Despite this assumption of local equilibrium, we will show that autonomous chemostats and reaction topology can lead to the emergence of genuine nonreciprocity.
Linear irreversible thermodynamics allows us to express the relative spatial flux, $\mathbf{J}^\alpha$, in terms of gradients of the chemical potential:
    \begin{align}
        \mathbf{J}^\alpha =& -\overline{\mathbf{L}} \cdot \partial_\alpha \boldsymbol{\mu}, \\ 
        \boldsymbol{\mu} \equiv& \frac{\delta F}{\delta \boldsymbol{\rho}}. \label{eq:constJs}
    \end{align}
The Onsager matrix, $\overline{\mathbf{L}}$, contains scalar elements (for isotropic systems with spatial parity symmetry) that depend on $\boldsymbol{\rho}$.
This matrix must be symmetric due to Onsager reciprocity and, in accordance with the second law, must also be positive-definite (PD)~\cite{Onsager1931A, Onsager1931B, degroot2013}.
We have introduced the vector of chemical potentials of the evolving species, $\boldsymbol{\mu}$, which is the functional derivative of the free energy, $F$.
In the absence of chemical reactions, this form of the spatial flux ensures that $F$ is a Lyapunov functional for these dynamics.
We note that as our chemostats are \textit{ideal}, they do not impact the chemical potentials of our $n_e$ dynamically evolving species.

In addition to driving fluxes, the local equilibrium hypothesis envisions that $\boldsymbol{\mu}$ also serves as the thermodynamic driving force for chemical reactions~\cite{Bazant2013, weber2019physics,Kirschbaum2021, bauermann2022energy, Aslyamov2023, Avanzini2021}.
Assuming reactions are rare events, the rates take the form:
    \begin{equation}
        \label{eq:jr}
        j_r^\pm = k_r \exp \left( \beta \sum_{i=1}^{n_e} \nu_{ri}^{\pm} \mu_i + \beta \sum_{i=1}^{n_c} \nu_{ri}^{\pm, \rm chemo} \mu_i^{\rm chemo} \right),
    \end{equation}
where $\beta = 1/k_{\rm B}T$ is the inverse thermal energy, $\boldsymbol{\mu}^{\rm chemo}$ is a length $n_c$ vector containing the chemical potential of the chemostats, and $k_r$ is the rate coefficient of reaction $r$ (which subsumes both a dynamic pre-exponential factor and the transition state free energy barrier) that respects the de Donder relation~\cite{DeDonder1927}.
For simplicity, we assume each $k_r$ is independent of $\boldsymbol{\rho}$~\footnote{This corresponds to assuming $n$-body reactions are purely $n$-body events that are unaffected by other degrees of freedom; the term resulting from a dependence of $k_r$ on $\boldsymbol{\rho}$ does not enter the linearized dynamics near DB steady states but does near CB and non-CB steady states.}.
\end{subequations}
In the absence of chemostats, the system is closed and $F$ remains a Lyapunov functional.
In open systems, however, the identification of a Lyapunov functional is not guaranteed as chemostats can drive the system to nonequilibrium steady states~\cite{Note2}.
Even for open systems near thermodynamic equilibrium, the Lyapunov functional cannot be immediately identified but can be derived as demonstrated by Ref.~\cite{Avanzini2024}.

The dynamics presented above are nonlinear with respect to the species concentrations and are anticipated to hold when the system remains in a state of local equilibrium.
We now consider the dynamics near a spatially uniform steady state solution to Eq.~\eqref{eq:gendyn}.
The spatially constant species densities are defined as $\boldsymbol{\rho}^{\rm SS}$ and the chemical potentials follow as $\boldsymbol{\mu}^{\rm SS} \equiv \boldsymbol{\mu}(\boldsymbol{\rho}^{\rm SS})$. 
The departure of the concentrations from their steady-state value is defined as $\delta \boldsymbol{\rho} \equiv \boldsymbol{\rho} - \boldsymbol{\rho}^{\rm SS}$. 
We can linearize the species dynamics [Eq.~\eqref{eq:gendyn}] with respect to $\delta \boldsymbol{\rho}$:
\begin{subequations}
\begin{equation}
\label{eq:lindynamics}
    \partial_t \delta \tilde{\boldsymbol{\rho}}_q \approx -\left( \boldsymbol{\mathcal{V}} + \overline{\mathbf{L}} q^2  \right) \cdot \mathbf{H}(q) \cdot \delta \tilde{\boldsymbol{\rho}}_q = \mathbf{A}(q) \cdot \delta \tilde{\boldsymbol{\rho}}_q,
\end{equation}
where $ \delta \tilde{\boldsymbol{\rho}}_q$ is the spatial Fourier transform of $\delta \boldsymbol{\rho}$ at wavenumber $q$ and $\mathbf{H}(q)$ is the (symmetric) thermodynamic Hessian of the free energy $F$ which can be expanded in even powers of $q$ as $\mathbf{H}=\mathbf{H}_0 + \mathbf{H}_2 q^2 + \mathbf{H}_4 q^4 + \cdots$~\footnote{If this series is truncated at order $n$, $\mathbf{H}_n$ must be PD to ensure infinitesimally small fluctuations (i.e.,~$\delta \tilde{\boldsymbol{\rho}}_{q}$ in the limit $q \rightarrow \infty$) are penalized.}.
The components of the $n_e \times n_e$ matrix $\boldsymbol{\mathcal{V}}$ take the form:
\begin{equation}
    \label{eq:Vdef}
    \mathcal{V}_{ij} \equiv \beta \sum_{r=1}^{n_r} \nu_{ir} \left( j_r^{-, \rm SS} \nu_{rj}^- - j_r^{+, \rm SS} \nu_{rj}^+ \right),
\end{equation}
where $j_r^{\pm, \rm SS} \equiv j_r^\pm(\boldsymbol{\rho}^{\rm SS})$.
All coefficients appearing in Eq.~\eqref{eq:lindynamics} are evaluated at the steady state of interest. 
The Jacobian can now be compactly expressed as:
    \begin{equation}
        \mathbf{A}(q) \equiv -\mathbf{L}(q) \cdot \mathbf{H}(q),
    \end{equation}
    where we have defined the generalized Onsager matrix:
    \begin{equation}
        \mathbf{L}(q) \equiv  \boldsymbol{\mathcal{V}} + \overline{\mathbf{L}} q^2,
    \end{equation}
    which now includes the effects of chemical reactions in addition to spatial transport.
\end{subequations}

The decomposition of our Jacobian into the product of two matrices, the symmetric PD Hessian and our generalized Onsager matrix, allows us to readily identify the criteria for achieving genuine nonreciprocity (i.e.,~achieving complex eigenvalues of $\mathbf{A}$).
As detailed in Appendix~A, the symmetry and definiteness of $\mathbf{H}$ require that $\mathbf{L}$ be asymmetric (or indefinite) in order for the Jacobian to have complex eigenvalues.
As $\overline{\mathbf{L}}q^2$ is symmetric PSD, genuine nonreciprocity thus requires that $\boldsymbol{\mathcal{V}}$ is asymmetric or indefinite.
We therefore turn our attention to the structure of $\boldsymbol{\mathcal{V}}$ which clearly depends on both the rates and the topology of the reaction networks under consideration.
It is thus useful to introduce some concepts in reaction network topology. 
We first define the two complexes associated with reaction $r$ as $\gamma_r^\pm \equiv \sum_{i=1}^{n_e} \nu_{ri}^\pm Z_i$ and the total number of unique complexes to be $n_\gamma$.
The forward and backward stoichiometric matrices may then be factorized as $\boldsymbol{\nu}^\pm =  \boldsymbol{\Delta}^\pm \cdot \boldsymbol{\Gamma}$, where $\boldsymbol{\Delta}^\pm$ is a $n_r \times n_\gamma$ matrix whose elements $\Delta^\pm_{r \gamma}$ are one if $\gamma=\gamma_r^\pm$ and zero otherwise while $\boldsymbol{\Gamma}$ is an $n_\gamma \times n_e$ matrix that encodes the stoichiometry of each species in each complex.
The incidence matrix of the directed graph of complexes follows as $\boldsymbol{\Delta} \equiv \boldsymbol{\Delta}^- - \boldsymbol{\Delta}^+$.

The matrix $\boldsymbol{\mathcal{V}}$ also depends explicitly on the steady-state reaction rates. 
The stoichiometric matrices and graph of complexes allow us to further distinguish between different classes of generation-free steady states -- steady states in which there is no net consumption/production of any species, i.e.,~$\mathbf{j}^{\rm SS} \cdot \boldsymbol{\nu} = \mathbf{0}$ where $\mathbf{j}^{\rm SS} \equiv \mathbf{j}^{+, \rm SS} - \mathbf{j}^{-, \rm SS}$. 
The steady states can be classified as~\cite{Horn1972, Feinberg1972, Avanzini2021, Avanzini2024}:
\begin{enumerate}
    \item \textit{detailed-balanced} (DB) when $\mathbf{j}^{\rm SS}=\mathbf{0}$,
    \item \textit{complex-balanced} (CB) when $\mathbf{j}^{\rm SS}\neq \mathbf{0}$ but $\mathbf{j}^{\rm SS} \cdot \boldsymbol{\Delta} = \mathbf{0}$,
    \item \textit{non-complex-balanced} when $\mathbf{j}^{\rm SS} \cdot \boldsymbol{\Delta}\neq \mathbf{0}$ but $\mathbf{j}^{\rm SS} \cdot \boldsymbol{\nu} = \mathbf{0}$.
\end{enumerate}
For closed systems -- or, more generally, for reaction networks where $\boldsymbol{\nu}$ has no left null vectors (a topological property) -- steady states are necessarily DB.
As $j_r^{+, \rm SS} = j_r^{-, \rm SS}$ in these steady states, the components of $\boldsymbol{\mathcal{V}}$ take the form:
\begin{equation}
    \mathcal{V}_{ij} = \beta \sum_r^{n_r} \nu_{ir} \nu_{rj} j_r^{\pm, \rm SS},
\end{equation}
and thus $\boldsymbol{\mathcal{V}}$ is symmetric PSD, preserving Onsager reciprocity.
The recovery of Onsager symmetry near DB steady states prevents the emergence of nonreciprocity in $\mathbf{A}$.
Moreover, as $\mathbf{L}$ is also PD, the stability of the system is entirely controlled by the Hessian, consistent with our thermodynamic expectations at or near equilibrium~\cite{Note2}. 
Finally, near thermodynamically stable (i.e.,~$\mathbf{H}$ is PD) DB steady states, this structure of $\mathbf{L}$ ensures that the linearized dynamics are are always \textit{coercive}~\cite{villani2009hypocoercivity, Note2}: small fluctuations from this steady state experience strict exponential decay.

We now turn our attention to the structure of $\boldsymbol{\mathcal{V}}$ in non-DB steady states -- these states are enabled by the left null space of $\boldsymbol{\nu}$.
When $\boldsymbol{\nu}$ and $\boldsymbol{\Delta}$ share the same null space (a topological property of the reaction network known as ``zero deficiency'') or, equivalently, when $\boldsymbol{\Gamma}$ has no left null space, non-DB steady states must be CB~\cite{Rao2016}.
In Appendix~B, we demonstrate that $\boldsymbol{\mathcal{V}}$ \textit{is generally asymmetric} in CB steady states. 
Crucially, the symmetric part of $\boldsymbol{\mathcal{V}}$ remains PSD (in non-CB steady states $\boldsymbol{\mathcal{V}}$ is generally asymmetric with an indefinite symmetric part).
CB steady states thus allow $\boldsymbol{\mathcal{V}}$ to be asymmetric and, as a result, Onsager reciprocity is broken at the level of species generation.
This broken Onsager reciprocity can lead to \textit{genuine nonreciprocity} (i.e.,~complex eigenvalues) in the Jacobian and give rise to oscillations.
However, despite this lack of reciprocity, the linearized dynamics near thermodynamically stable CB steady states are always coercive~\cite{Note2, villani2009hypocoercivity}, just as was the case for DB steady states.

Intriguingly, Avanzini~\textit{et al.}~\cite{Avanzini2024} recently found that, despite being out of equilibrium, reaction networks that have a CB steady state still have a Lyapunov functional when the chemostats are ideal~\footnote{This was generalized to non-ideal chemostats in Ref.~\cite{Avanzini2024} by introducing auxilliary, ideal chemostats that the the non-ideal chemostats equilibrate with on fast time scales relative to other reactions.}.
In particular, for a CB steady state with chemical potentials $\boldsymbol{\mu}^{\rm SS}$, they found that the following grand potential acts as a Lypanuov functional:
\begin{equation}
    \label{eq:grandpotential}
    \Omega \equiv F - \boldsymbol{\mu}^{\rm SS} \cdot \int_V d \mathbf{x} \boldsymbol{\rho}(\mathbf{x}) - \boldsymbol{\mu}^{\rm chemo} \cdot \mathbf{N}^{\rm chemo},
\end{equation}
where $\mathbf{x}$ specifies spatial position and $\mathbf{N}^{\rm chemo}$ is a length $n_c$ vector containing the particle number of each chemostatted species.
Importantly, Avanzini~\textit{et al.}~\cite{Avanzini2024} refer to this grand potential as a ``kinetic potential'' as $\boldsymbol{\mu}^{\rm SS}$ are the chemical potentials at the nonequilibrium steady state.
We emphasize that the steady-state concentrations the system relaxes to are consistent with those obtained via the minimization of this grand potential.
For the ideal chemostats considered here, each element of $\mathbf{N}^{\rm chemo}$ is a constant and the last term in Eq.~\eqref{eq:grandpotential} is therefore inconsequential.
In the SM~\cite{Note2}, we demonstrate how the symmetry and kernel (i.e.,~null space) of $\boldsymbol{\mathcal{V}}$ determines the existence and form of the Lyapunov functional near a steady state.
This makes clear that CB steady states can have oscillatory instabilities which necessarily saturate to a stationary state.
Furthermore, the species concentrations at this steady state coincide with thermodynamic equilibrium in a fully open ensemble.

\begin{figure}
    \centering
    \includegraphics[width=0.99\linewidth]{reaction_stoichiometry_chemostatting_schematic2.png}
    \caption{Schematic of the steady state in an open system whose species interconvert through a cyclic reaction network of length $n=3$. The reactions are $A \rightleftharpoons B + D$, $B \rightleftharpoons C + E$, and $C \rightleftharpoons A + G$ where $D$, $E$, and $G$ are chemostatted species which are exchanged with an external reservoir held at fixed chemical potential. The length of the arrows reflects the steady-state reaction rate in a particular direction; here $j_r^{+, \rm SS} > j_r^{-, \rm SS}~\forall \ r$ as we take $\mu_i^{\rm chemo}=\mu^{\rm chemo} < 0$ for $i \in \{D, E, G\}$. As a result, the chemostatted species are continuously output into the reservoir. We use the simplified notation $j_r^{\pm, \rm SS} \rightarrow j^\pm$ in the text within the schematic. The chemostats thus bias the internal cycle in the clockwise direction such that the only steady state is CB -- in closed systems, the reaction rates (arrow lengths) are equal in both directions for all reactions.}
    \label{fig:stoich}
\end{figure}

\textit{Nonreciprocity in Cyclic Reaction Networks.--} We will now demonstrate that complex-balancing provides a generic route to oscillatory instabilities while retaining a Lyapunov functional.
Consider an open reaction network containing $n$ reactions, the $i$th of which takes the form ${Z_i \rightleftharpoons Z_{i + 1} + Z_{i+1}^{\rm chemo}}$. 
We consider \textit{cyclic} reaction networks where the $n$th reaction completes the cycle,~${Z_{n} \rightleftharpoons Z_{1} + Z_{1}^{\rm chemo}}$ (i.e.,~index $i+n$ corresponds to index $i$), and thus ${n = n_r = n_e = n_c}$.
In Fig.~\ref{fig:stoich}, we illustrate the form of this cyclic network for $n=3$.

Consider the limit $\mu_i^{\rm chemo} \rightarrow - \infty$ for each of the chemostatted species, which alters the form of the rates described in Eq.~\eqref{eq:jr}.
We now have $j_r^- = 0~\forall \ r$, while $j_r^+$ takes the form:
\begin{equation}
    \label{eq:jr+example}
    j_r^+ = k_r \exp \left( \beta \mu_r \right),
\end{equation}
where $\mu_r$ is the chemical potential of the reactant species in reaction $r$.
In this limit, the generation terms in Eq.~\eqref{eq:gendyn1} take the form ${\sum_r^{n}j_r \nu_{ri} = j^+_{i-1} - j_i^+}$ (with ${j^+_{0}=j^+_{n}}$).
As a result, the only possible steady states in which there is no species generation are CB.
In these steady states, there is an identical reaction rate across every reaction in the cycle, with ${j^{+, \rm SS}_r = j^{\rm SS} > 0}$.
This corresponds to a scenario in which the following effective chemical potential ${\mu_r + k_B T \log k_r}$ is equal between each reaction, with the contribution from $k_r$ reflecting a distinctly nonequilibrium effect.

Before choosing a model for $\boldsymbol{\mu}$ and linearizing about a CB steady state, we note that $\boldsymbol{\mathcal{V}}$ must take the following form:
\begin{equation}
    \boldsymbol{\mathcal{V}} = \beta j^{\rm SS} \left( \mathbf{I} - \mathbf{C}^{(n)} \right),
\end{equation}
where $C^{(n)}_{ij}=\delta_{(i-1)j} + \delta_{i1} \delta_{n j}$ is the $ij$ component of the ${n \times n}$ cyclic lower shift matrix.
For these biased cyclic reactions, $\boldsymbol{\mathcal{V}}$ is thus \textit{always} asymmetric.

\begin{figure}
    \centering
    \includegraphics[width=0.99\linewidth]{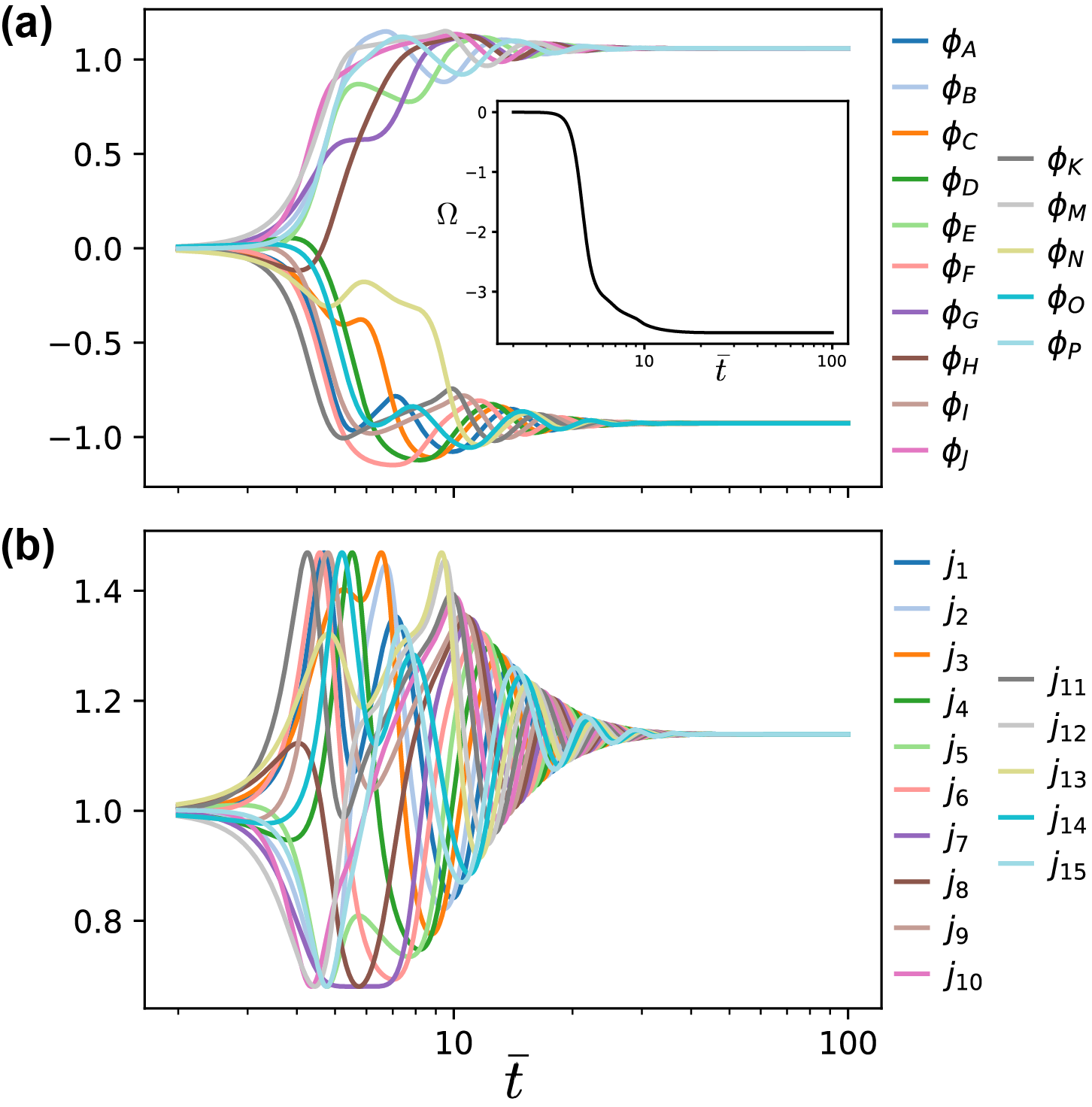}
    \caption{Time evolution of (a) the dimensionless relative densities and (b) reaction rates (in units of $k$) in a cyclic network of $n=15$ reactions with CB steady states, found by simulating the full nonlinear dynamics in Eq.~\eqref{eq:phidt}. Inset in (a) displays the evolution of the grand potential $\Omega$.}
    \label{fig:hopf}
\end{figure}

The structure of $\boldsymbol{\mathcal{V}}$ for biased cyclic reaction networks allows for an oscillatory dynamical response while $\Omega$ [Eq.~\eqref{eq:grandpotential}] is anticipated to be a Lyapunov functional. Whether oscillations are actually observed depends also on the form of the thermodynamic Hessian.
As our aim is to numerically corroborate emergent nonreciprocity while retaining a Lyapunov functional, we focus on a simple model system. 
We again consider spatially uniform steady states and, for simplicity, will neglect spatial gradients in the perturbed response.
The dynamics no longer depend on space (i.e.,~$\boldsymbol{\rho}$ can only change globally) and, correspondingly, the Jacobian now takes the form ${\mathbf{A} =-\boldsymbol{\mathcal{V}}\cdot \mathbf{H}}$ which is independent of $q$.
To further simplify, we assume $k_r = k~\forall \ r$ and, to accentuate the role of $\boldsymbol{\mathcal{V}}$, use chemical potentials of the form ${\beta \mu_i = -\phi_i + \phi_i^3}$ where we have defined the dimensionless relative density $\boldsymbol{\phi} \equiv v (\boldsymbol{\rho} - \boldsymbol{\rho}^{\rm ref})$ where $\boldsymbol{\rho}^{\rm ref}$ is a vector of reference densities and $v$ is a characteristic volume.
This form of $\mu_i$ will result in instabilities at $\phi_i=0$.
The full nonlinear dynamics can then be expressed as:
\begin{equation}
    \label{eq:phidt}
    \frac{d \phi_i}{d \bar{t}} = \exp\left( -\beta \mu_{i-1}\right) - \exp\left(- \beta \mu_i\right),
\end{equation}
where $\bar{t} \equiv t / \tau$ is a dimensionless time with $\tau \equiv (k v)^{-1}$.
As this reaction network is cyclic, we again emphasize our use of periodic boundary conditions (i.e.,~${\mu_0=\mu_n}$).

We now focus on the CB steady state with $\boldsymbol{\phi} = \mathbf{0}$, resulting in $j^{\rm SS} = k$. 
For this choice of chemical potential, the steady-state Hessian is diagonal, ${\beta \mathbf{H} = -v\mathbf{I}}$. 
The linearized dynamics of our density perturbations near this steady state follow as:
\begin{equation}
    \frac{ d \boldsymbol{\phi}}{d \bar{t}} \approx \left( \mathbf{I} - \mathbf{C}^{(n)} \right) \cdot \boldsymbol{\phi},
\end{equation}
where we identify the Jacobian as $\mathbf{A} = j^{\rm SS} v ( \mathbf{I} - \mathbf{C}^{(n)} )$. 
The linear stability of this steady state is encoded in the $n$ eigenvalues of $\mathbf{A}$. 
Since $\mathbf{A}$ is circulant, its spectrum can be obtained analytically: the $m$th eigenvalue is ${\lambda_m = j^{\rm SS} v (1 - e^{-2\pi i m/n})}$ for ${m = 0, 1, \cdots, n-1 }$.
The $m=0$ mode is marginal with $\lambda_0=0$ [reflecting the time-independence of $\phi^{\rm tot} \equiv \sum_i^{n_e} \phi_i$, as expected from Eq.~\eqref{eq:phidt}], whereas the other $n-1$ modes are unstable (i.e.,~${\Re(\lambda_m)>0~\forall \ m>0}$) and have imaginary components that will result in an oscillatory response.
We can estimate the relative magnitude of oscillations in these unstable modes by comparing the imaginary and real components of $\lambda_m$:
\begin{equation}
    \frac{\Im(\lambda_m)}{\Re(\lambda_m)} = \cot\left(\frac{\pi m} {n}\right) \approx \frac{n}{m \pi},
\end{equation}
where we have expanded about large $n/m$.
We can see that increasing $n$ amplifies the relative magnitude of oscillations and we therefore consider systems with $n > 3$.

Figure~\ref{fig:hopf}a displays the trajectory $\boldsymbol{\phi}(t)$ determined from direct numerical integration of Eq.~\eqref{eq:phidt} with weak initial noise added to each $\phi_i$ for $n=15$.
The species concentrations undergo significant temporal oscillations before saturating to a new CB steady state while the free energy, $\Omega$, monotonically decreases (see the inset of Fig.~\ref{fig:hopf}a).
The concentrations are truly time-independent in this new steady state and do not enter into a limit cycle with sustained oscillations.
The absence of a limit cycle, which are often observed in nonreciprocal systems~\cite{Saha2020, You2020, Fruchart2021, Brauns2024}, is a reflection that our dynamics indeed have a Lyapunov functional.
Intriguingly, the \textit{lower free energy} CB steady state the system reaches has \textit{higher reaction rates} (see Fig.~\ref{fig:hopf}b).
This high-rate steady state can be directly predicted by minimizing $\Omega$, subject to the constraint of fixed $\phi^{\rm tot}$, as we demonstrate for $n=3$ in the SM~\cite{Note2}.

Interestingly, we can alter the reaction network topology by ``cutting'' the cycle to create a biased reaction chain.
This amounts to eliminating our periodic boundary conditions which prevents the possibility of CB steady states and, consequently, the possibility of genuine nonreciprocity.
The final DB steady state for these biased reaction chains will result in the exponential localization of species concentrations at the biased boundary.
In fact, if one uses ideal chemical potentials for each species, one precisely recovers a classical Hatano-Nelson model~\cite{Hatano1996, Hatano1997, Zhang2022} used to describe non-Hermitian skin effects. 
The genuine nonreciprocity found under periodic boundary conditions (which causes the eigenvalues of $\mathbf{A}$ to trace a circle in the complex plane) results in skin effects upon cutting the cycle, consistent with the arguments put forth in Refs.~\cite{Yokomizo2019,Zhang2020}.
We describe these chemical skin effects in further detail in the SM~\cite{Note2}.

\textit{Discussion and Conclusions.--} 
In equilibrium, asymmetry of the Jacobian is innocuous: it is always isospectral to a symmetric Jacobian, precluding temporal oscillations.
Here we show how this constraint is absent in open chemical reaction networks near a complex-balanced steady state, where temporal oscillations of species concentrations coincide with a monotonically decreasing free energy.
The origins of temporal oscillations in chemical reaction networks near complex-balanced steady states are rooted in the nonequilibrium driving provided by the chemostats.
Nevertheless, the local equilibrium hypothesis -- which posits that the reaction rates and diffusion remain driven by thermodynamic forces -- preserves the existence of a Lypanuov functional in the form of the grand potential.

The preservation of thermodynamic driving forces with nonequilibrium effects that generate \textit{transverse} forces is analogous to a particle moving on a potential energy landscape according to overdamped Langevin dynamics with spatially odd drag.
Such dynamics are similar to those of a tracer particle in a chiral fluid, for example~\cite{poggioli2023}. 
The particle moves down potential energy gradients, lowering the energy, with odd drag resulting in spatial oscillations on the way to an energy minimum.
This analogy is further described in the SM.
In both scenarios, the systems are driven out of equilibrium in a manner that allows a Lyapunov functional to still exist.
In fact, the transverse forces induced by the odd drag can accelerate the dynamics while retaining the equilibrium steady state as a solution~\cite{hwang1993accelerating, hwang2005accelerating, Ohzeki2015, kaiser2017acceleration, Gao2020, futami2020, Ghimenti2022, Ghimenti2023, Ghimenti2024a, Ghimenti2024b}.
This nonequilibrium speedup is achieved by mixing the system out of slow parts of the dynamics -- our work may offer an alternative \textit{chemical} platform to realize this in the dynamics of large, disordered reaction networks.
Hypocoercivity~\cite{villani2009hypocoercivity} is an extreme example of this mixing-induced speedup, where oscillations kick the system out of the kernel (i.e.,~null space) of the gradient part of the dynamics which allows relaxation to proceed.
In fact, the dynamics are hypocoercive near some non-CB steady states, allowing the notion of oscillations in the presence of a Lyapunov functional to be extended to these systems (see Appendix~C for an extended discussion).
Our findings demonstrate that thermodynamic forces in chemical reaction networks can drive transitions between steady states with transverse nonequilibrium forces generating temporal oscillations during these transitions. 

\begin{acknowledgments}
We thank Cheyne Weis for insightful discussions and Yu-Jen Chiu and Eric Weiner for helpful feedback on this manuscript. 
D.E. acknowledges partial support from the U.S. Department of Defense through the National Defense Science and Engineering Graduate Fellowship Program.
\end{acknowledgments}

\newpage

\section{Appendices}

\textit{Appendix A: nonreciprocity in equilibrium.--} We now demonstrate that, even at equilibrium, the Jacobian is generally asymmetric; however, its eigenvalues must be purely real.
In particular, we consider systems where the steady state is DB and thus $\boldsymbol{\mathcal{V}}$ is symmetric PSD.
The Jacobian $\mathbf{A}^{\rm eqm}(q) = -\mathbf{L}^{\rm eqm}(q) \cdot \mathbf{H}(q)$ is the negative of the product of a symmetric PD $\mathbf{L}^{\rm eqm}(q)$ and the symmetric $\mathbf{H}(q)$.
Choosing random $\mathbf{L}^{\rm eqm}$ and $\mathbf{H}$ (with $n_e>1$) typically yields an asymmetric $\mathbf{A}^{\rm eqm}$ whose spectrum remains strictly real.
This is because $\mathbf{A}^{\rm eqm}$ is always similar to a symmetric Jacobian in equilibrium and similarity preserves the spectrum.
To see this, first note that as $\mathbf{L}^{\rm eqm}$ is symmetric PD for $q>0$, it admits the factorization:
\begin{equation}
    \mathbf{L}^{\rm eqm} = \sqrt{\mathbf{L}^{\rm eqm}} \cdot \sqrt{\mathbf{L}^{\rm eqm}},
\end{equation}
where $\sqrt{\mathbf{L}^{\rm eqm}}$ is itself symmetric PD.
As $\mathbf{H}$ is symmetric, one can then define the \textit{symmetric} matrix:
\begin{align}
    \overline{\mathbf{A}}^{\rm eqm} &\equiv \sqrt{\mathbf{L}^{\rm eqm}}^{-1} \cdot \mathbf{A}^{\rm eqm} \cdot \sqrt{\mathbf{L}^{\rm eqm}} \nonumber \\  &= -\sqrt{\mathbf{L}^{\rm eqm}} \cdot \mathbf{H} \cdot \sqrt{\mathbf{L}^{\rm eqm}},
\end{align}
which is similar to $\mathbf{A}^{\rm eqm}$.
Indeed, this is because $\mathbf{L}^{-1}$ plays the role of a Riemannian metric with respect to which the linearized dynamics correspond to gradient descent and the Jacobian is accordingly symmetric.
In other words, passive systems generally do not relax to thermodynamic equilibrium along the path of steepest descent of the free energy unless $\mathbf{L}^{\rm eqm}$ is proportional to the identity matrix.
Consequently, the eigenvalues of $\mathbf{A}^{\rm eqm}$ must be purely real, precluding instabilities with a temporally oscillating character, which require complex eigenvalues~\cite{Hohenberg1977, Frohoff20232, Greve2024}.
When $q=0$ and $\mathbf{L}^{\rm eqm}$ is PSD, the above results apply on the space of species concentrations $\{ \boldsymbol{\rho} \in \mathbb{R}^{n_e} | \boldsymbol{\rho} \geq \mathbf{0} \}$ modulo the right null space of $\mathbf{L}^{\rm eqm}$.
Indeed, as perturbations along this null space are associated with changes in the values of the conserved quantities, these perturbations do not relax and instead correspond to changes in the global conditions the system is held under.
In non-DB steady states where $\mathbf{L}^{-1}$ is asymmetric, these arguments no longer apply and $\mathbf{L}^{-1}$ cannot be treated as a (semi-)Riemannian metric~\footnote{This contrasts \textit{spurious} gradient dynamics (observed in some models with nonreciprocal $\mathbf{H}$~\cite{Frohoff2023, Greve2025}) where $\mathbf{L}$ and $\mathbf{H}$ can be redefined such that both are symmetric but indefinite so that $\mathbf{L}^{-1}$ gives a semi-Riemannian metric underlying a spurious gradient flow. This can occur in non-CB steady states if $\boldsymbol{\mathcal{V}}$ is symmetric but indefinite.}.

Beyond being purely real, $\mathbf{A}^{\rm eqm}$ has the same numbers of positive, negative, and zero eigenvalues as $\mathbf{H}$ (formally, $\mathbf{A}^{\rm eqm}$ inherits the inertia of $\mathbf{H}$).
This follows from Sylvester's Law of Inertia~\cite{Sylvester1852}: $\mathbf{H}$ is congruent to $\overline{\mathbf{A}}^{\rm eqm}$ and $\overline{\mathbf{A}}^{\rm eqm}$ is similar to  $\mathbf{A}^{\rm eqm}$. 
As a result, equilibrium instabilities are necessarily rooted in negative free-energy curvature (all instabilities in DB are thus ``E-type'' according to the classification in Ref.~\cite{Aslyamov2023}). 
Furthermore, if the free energy takes a Cahn-Hilliard~\cite{Cahn1958} form where $\mathbf{H}_2$ is PD and $\mathbf{H}_n=\mathbf{0}$ for $n>2$, the eigenvalues of $\mathbf{H}(q)$ increase strictly with $q$ since $\mathbf{H}_2$ is PD.
In these systems, stability in the limit $q \rightarrow 0$ therefore guarantees stability for $q > 0$.
This rules out small-scale (e.g.,~Turing~\cite{Turing1952}) instabilities~\cite{Frohoff20232, Greve2024} in equilibrium systems that are well-described by Cahn-Hilliard free energies.
Similar statements can be made about a restriction to ``E''-type instabilities, and a lack of small-scale instabilities in systems described by Cahn-Hilliard free energies, in CB steady states.
When an instability occurs, the real part of an eigenvalue of $\mathbf{A}$ crosses zero, changing the sign of $\det (\mathbf{A}) = \det (\mathbf{L}) \det (\mathbf{H})$.
As $\det (\mathbf{L}) \geq 0$ in CB steady states (since the symmetric part of $\mathbf{L}$ is PSD), an eigenvalue of $\mathbf{A}$ therefore only changes sign if an eigenvalue of $\mathbf{H}$ does.

\textit{Appendix B: $\boldsymbol{\mathcal{V}}$ in complex-balanced steady states.--} It is convenient to express $\boldsymbol{\mathcal{V}}$ as $\boldsymbol{\mathcal{V}} = \boldsymbol{\Gamma}^{\top} \cdot \boldsymbol{\mathcal{V}}^{\rm complex} \cdot \boldsymbol{\Gamma}$, where the $\gamma \gamma'$ element of $\boldsymbol{\mathcal{V}}^{\rm complex}$ is: 
\begin{equation}
    \label{eq:Vgammadef}
    \mathcal{V}^{\rm complex}_{\gamma \gamma'} \equiv \beta\sum_{r=1}^{n_r} \Delta_{\gamma r} \left( j_r^{-, \rm SS} \Delta_{r \gamma'}^- - j_r^{+, \rm SS} \Delta_{r \gamma'}^+ \right).
\end{equation}
The diagonal elements $\gamma=\gamma'$ will always be positive since $\Delta_{\gamma r}=0$ if complex $\gamma$ is not involved in reaction $r$, $\Delta_{\gamma r}=1$ if complex $\gamma$ is a product, and $\Delta_{\gamma r}=-1$ if complex $\gamma$ is a reactant.
In particular, 
\begin{multline}
    \mathcal{V}_{\gamma \gamma'}^{\rm complex} \equiv \beta \delta_{\gamma \gamma'} \sum_r^{n_r} \left( \delta_{\gamma R_r} j_r^{+, \rm SS} + j_r^{-, \rm SS} \delta_{\gamma P_r} \right) \\ - \beta \left( \sum_{r \in \{ \gamma \rightarrow \gamma' \}} j_r^{-, \rm SS} + \sum_{r \in \{ \gamma' \rightarrow \gamma \}} j_r^{+, \rm SS} \right),
\end{multline}
where $\delta_{\gamma R_r}$ equals $1$ if $\gamma$ is the reactant of reaction $r$ and $0$ otherwise, while $\delta_{\gamma P_r}$ equals $1$ if $\gamma$ is the product of reaction $r$ and $0$ otherwise.
Here, $\{ \gamma \rightarrow \gamma' \}$ denotes the set of all reactions which convert complex $\gamma$ into complex $\gamma'$.
Using the CB condition:
\begin{equation}
    \sum_r^{n_r} \left(\delta_{\gamma R_r} j_r^{+, \rm SS} + \delta_{\gamma P_r} j_r^{-, \rm SS}\right) = \sum_r^{n_r} \left(\delta_{\gamma P_r} j_r^{+, \rm SS} + \delta_{\gamma R_r} j_r^{-, \rm SS}\right),
\end{equation}
we find:
\begin{equation}
    \sum_{\gamma' \neq \gamma} \left| \mathcal{V}_{\gamma \gamma'}^{\rm complex} \right| = \mathcal{V}_{\gamma \gamma}^{\rm complex}.
\end{equation}
Thus, Gershgorin's circle theorem~\cite{Varga2011} implies that the real part of every eigenvalue of $\boldsymbol{\mathcal{V}}^{\rm complex}$ must be positive.
Moreover, we notice that $\boldsymbol{\mathcal{V}}^{\rm complex}$ has the structure of a weighted graph Laplacian~\cite{Chung2005}, albeit a directed (asymmetric) one as the steady-state rate at which $\gamma$ is converted into $\gamma'$ is not necessarily the same as the steady-state rate at which $\gamma'$ is converted into $\gamma$ (except for in DB).

The components of the symmetric part of $\boldsymbol{\mathcal{V}}^{\rm complex}$, $\boldsymbol{\mathcal{V}}^{\rm complex, S}$, are:
\begin{multline}
    \mathcal{V}_{\gamma \gamma'}^{\rm complex, S} = \beta \delta_{\gamma \gamma'} \sum_r^{n_r} \left( \delta_{\gamma R_r} j_r^{+, \rm SS} + j_r^{-, \rm SS} \delta_{\gamma P_r} \right) \\ - \frac{\beta}{2} \left[ \sum_{r \in \{ \gamma \rightarrow \gamma' \}} \left( j_r^{-, \rm SS} + j_r^{+, \rm SS} \right) + \sum_{r \in \{ \gamma' \rightarrow \gamma \}} \left( j_r^{-, \rm SS} + j_r^{+, \rm SS} \right) \right],
\end{multline}
and we again have:
\begin{equation}
    \sum_{\gamma' \neq \gamma} \left| \mathcal{V}_{\gamma \gamma'}^{\rm complex} \right| = \mathcal{V}_{\gamma \gamma}^{\rm complex}.
\end{equation}
It follows from Gershgorin's circle theorem that $\boldsymbol{\mathcal{V}}^{\rm complex, S}$ is PSD and therefore we have ${\mathbf{u} \cdot \boldsymbol{\mathcal{V}}^{\rm complex, S} \cdot \mathbf{u} \geq 0}$ for any vector $\mathbf{u}$.
The symmetric part of $\boldsymbol{\mathcal{V}}$, ${\boldsymbol{\mathcal{V}}^{\rm S} = \boldsymbol{\Gamma}^{\top} \cdot \boldsymbol{\mathcal{V}}^{\rm complex, S} \cdot \boldsymbol{\Gamma}}$, is then also PSD as for any vector $\mathbf{w}$, one can define $\mathbf{u} \equiv \boldsymbol{\Gamma} \cdot \mathbf{w}$ such that ${\mathbf{w} \cdot \boldsymbol{\mathcal{V}}^{\rm S} \cdot \mathbf{w} = \mathbf{u} \cdot \boldsymbol{\mathcal{V}}^{\rm complex, S} \cdot \mathbf{u} \geq 0}$.

\textit{Appendix C: hypocoercivity when non-complex-balanced.--} Consider the reaction network (we have already taken the limit of strong chemostatting in the forward direction which makes the reactions approximately unidirectional with $j_r^-\approx 0 \ \forall \ r$):
\begin{subequations}
    \begin{align}
        2A + 3B \rightarrow& A + 4B, \\
        B + 2C \rightarrow& 3C, \\
        A + 3B + 2C \rightarrow& 2A + 3B + C,
    \end{align}
\end{subequations}
which conserves the overall system concentration but not that of the individual species.
The above network cannot achieve a CB steady state but admits a \textit{non-CB steady state} when $j_r^+ = j^{\rm SS} \ \forall \ r$.
At this steady state, the matrix $\boldsymbol{\mathcal{V}}$ takes the form:
\begin{equation}
    \boldsymbol{\mathcal{V}} = \beta j^{\rm SS} \begin{bmatrix}
        1 & 0 & -2 \\ -2 & -2 & 2 \\ 1 & 2 & 0
    \end{bmatrix},
\end{equation}
which has a symmetric part that is \textit{indefinite} and therefore thermodynamically unstable states ($\mathbf{H}$ with negative eigenvalues) can be dynamically stable.
For example, when $\mathbf{H}$ is diagonal with $2$, $-1$, and $1$ along the diagonals, the Jacobian takes the form:
\begin{equation}
    \mathbf{A} = \begin{bmatrix}
        -2 & 0 & 2 \\ 4 & -2 & -2 \\ -2 & 2 & 0
    \end{bmatrix},
\end{equation}
which has eigenvalues $0$ (corresponding to the null vector $\begin{bmatrix} 1 & 1 & 1\end{bmatrix}^{\top}$) and $-2 \pm i 2 \sqrt{2}$.
Therefore, $\mathbf{A}$ is genuinely nonreciprocal.
The symmetric part of $\mathbf{A}$ is PSD with \textit{two} null vectors, $\begin{bmatrix} 1 & 1 & 0\end{bmatrix}^{\top}$ and $\begin{bmatrix} 0 & 0 & 1\end{bmatrix}^{\top}$ (which can be summed to achieve $\begin{bmatrix} 1 & 1 & 1\end{bmatrix}^{\top}$), whereas the antisymmetric part of $\mathbf{A}$ only has one null vector ($\begin{bmatrix} 1 & 1 & 1\end{bmatrix}^{\top}$).
Trajectories along $\begin{bmatrix} 1 & 1 & 0\end{bmatrix}^{\top}$ and $\begin{bmatrix} 0 & 0 & 1\end{bmatrix}^{\top}$ are mixed out of the kernel of the dissipative (symmetric part of $\mathbf{A}$) part of the dynamics by the antisymmetric part of $\mathbf{A}$.
In the absence of nonreciprocity, the dynamics would stall (i.e.,~trajectories will not move) in the kernel of the symmetric part of $\mathbf{A}$.
Instead, the transverse forces facilitate the exponential decay of fluctuations, which means the dynamics are hypocoercive~\cite{villani2009hypocoercivity}.
In other words, while the dissipative part of the dynamics alone is not strictly contracting since it leaves a nontrivial kernel, the oscillatory part can continually mix the system state out of the kernel and enable dissipation to act.
When this is the case, $-\int dq \delta \tilde{\boldsymbol{\rho}}_q \cdot \mathbf{A} \cdot \delta \tilde{\boldsymbol{\rho}}_q$ acts as a local Lyapunov functional for the linearized dynamics.
We therefore expect this quadratic Lyapunov functional to be minimized while concentrations experience oscillations, just as was the case for the systems described in the main text near CB steady states.
More broadly, temporal oscillations in species concentrations that decrease a quadratic Lyapunov functional occur whenever the linearized dynamics are hypocoercive: the system above represents one such case.


\begin{thebibliography}{85}%
\makeatletter
\providecommand \@ifxundefined [1]{%
 \@ifx{#1\undefined}
}%
\providecommand \@ifnum [1]{%
 \ifnum #1\expandafter \@firstoftwo
 \else \expandafter \@secondoftwo
 \fi
}%
\providecommand \@ifx [1]{%
 \ifx #1\expandafter \@firstoftwo
 \else \expandafter \@secondoftwo
 \fi
}%
\providecommand \natexlab [1]{#1}%
\providecommand \enquote  [1]{``#1''}%
\providecommand \bibnamefont  [1]{#1}%
\providecommand \bibfnamefont [1]{#1}%
\providecommand \citenamefont [1]{#1}%
\providecommand \href@noop [0]{\@secondoftwo}%
\providecommand \href [0]{\begingroup \@sanitize@url \@href}%
\providecommand \@href[1]{\@@startlink{#1}\@@href}%
\providecommand \@@href[1]{\endgroup#1\@@endlink}%
\providecommand \@sanitize@url [0]{\catcode `\\12\catcode `\$12\catcode `\&12\catcode `\#12\catcode `\^12\catcode `\_12\catcode `\%12\relax}%
\providecommand \@@startlink[1]{}%
\providecommand \@@endlink[0]{}%
\providecommand \url  [0]{\begingroup\@sanitize@url \@url }%
\providecommand \@url [1]{\endgroup\@href {#1}{\urlprefix }}%
\providecommand \urlprefix  [0]{URL }%
\providecommand \Eprint [0]{\href }%
\providecommand \doibase [0]{https://doi.org/}%
\providecommand \selectlanguage [0]{\@gobble}%
\providecommand \bibinfo  [0]{\@secondoftwo}%
\providecommand \bibfield  [0]{\@secondoftwo}%
\providecommand \translation [1]{[#1]}%
\providecommand \BibitemOpen [0]{}%
\providecommand \bibitemStop [0]{}%
\providecommand \bibitemNoStop [0]{.\EOS\space}%
\providecommand \EOS [0]{\spacefactor3000\relax}%
\providecommand \BibitemShut  [1]{\csname bibitem#1\endcsname}%
\let\auto@bib@innerbib\@empty
\bibitem [{\citenamefont {Berridge}\ \emph {et~al.}(1988)\citenamefont {Berridge}, \citenamefont {Cobbold}, \citenamefont {Cuthbertson}, \citenamefont {Downes},\ and\ \citenamefont {Hanley}}]{Berridge1988}%
  \BibitemOpen
  \bibfield  {author} {\bibinfo {author} {\bibfnamefont {M.~J.}\ \bibnamefont {Berridge}}, \bibinfo {author} {\bibfnamefont {P.~H.}\ \bibnamefont {Cobbold}}, \bibinfo {author} {\bibfnamefont {K.~S.~R.}\ \bibnamefont {Cuthbertson}}, \bibinfo {author} {\bibfnamefont {C.~P.}\ \bibnamefont {Downes}},\ and\ \bibinfo {author} {\bibfnamefont {M.~R.}\ \bibnamefont {Hanley}},\ }\bibfield  {title} {\bibinfo {title} {Spatial and temporal aspects of cell signalling},\ }\href {https://doi.org/10.1098/rstb.1988.0080} {\bibfield  {journal} {\bibinfo  {journal} {Philos. Trans. R. Soc. Lond. B Biol. Sci.}\ }\textbf {\bibinfo {volume} {320}},\ \bibinfo {pages} {325} (\bibinfo {year} {1988})}\BibitemShut {NoStop}%
\bibitem [{\citenamefont {Fewtrell}(1993)}]{Fewtrell1993}%
  \BibitemOpen
  \bibfield  {author} {\bibinfo {author} {\bibfnamefont {C.}~\bibnamefont {Fewtrell}},\ }\bibfield  {title} {\bibinfo {title} {Ca2+ oscillations in non-excitable cells},\ }\href {https://doi.org/https://doi.org/10.1146/annurev.ph.55.030193.002235} {\bibfield  {journal} {\bibinfo  {journal} {Annu. Rev. Physiol.}\ }\textbf {\bibinfo {volume} {55}},\ \bibinfo {pages} {427} (\bibinfo {year} {1993})}\BibitemShut {NoStop}%
\bibitem [{\citenamefont {Xiong}\ \emph {et~al.}(2025)\citenamefont {Xiong}, \citenamefont {Tong},\ and\ \citenamefont {Wu}}]{Xiong2025}%
  \BibitemOpen
  \bibfield  {author} {\bibinfo {author} {\bibfnamefont {D.}~\bibnamefont {Xiong}}, \bibinfo {author} {\bibfnamefont {C.~S.}\ \bibnamefont {Tong}},\ and\ \bibinfo {author} {\bibfnamefont {M.}~\bibnamefont {Wu}},\ }\bibfield  {title} {\bibinfo {title} {A molecular systems perspective on calcium oscillations beyond ion fluxes},\ }\href {https://doi.org/https://doi.org/10.1016/j.ceb.2025.102523} {\bibfield  {journal} {\bibinfo  {journal} {Curr. Opin. Cell Biol.}\ }\textbf {\bibinfo {volume} {94}},\ \bibinfo {pages} {102523} (\bibinfo {year} {2025})}\BibitemShut {NoStop}%
\bibitem [{\citenamefont {Raskin}\ and\ \citenamefont {De~Boer}(1999)}]{raskin1999rapid}%
  \BibitemOpen
  \bibfield  {author} {\bibinfo {author} {\bibfnamefont {D.~M.}\ \bibnamefont {Raskin}}\ and\ \bibinfo {author} {\bibfnamefont {P.~A.}\ \bibnamefont {De~Boer}},\ }\bibfield  {title} {\bibinfo {title} {Rapid pole-to-pole oscillation of a protein required for directing division to the middle of escherichia coli},\ }\href {https://doi.org/10.1073/pnas.96.9.4971} {\bibfield  {journal} {\bibinfo  {journal} {Proc. Nat. Acad. Sci.}\ }\textbf {\bibinfo {volume} {96}},\ \bibinfo {pages} {4971} (\bibinfo {year} {1999})}\BibitemShut {NoStop}%
\bibitem [{\citenamefont {Raskin}\ and\ \citenamefont {de~Boer}(1999)}]{raskin1999minde}%
  \BibitemOpen
  \bibfield  {author} {\bibinfo {author} {\bibfnamefont {D.~M.}\ \bibnamefont {Raskin}}\ and\ \bibinfo {author} {\bibfnamefont {P.~A.}\ \bibnamefont {de~Boer}},\ }\bibfield  {title} {\bibinfo {title} {Minde-dependent pole-to-pole oscillation of division inhibitor minc in escherichia coli},\ }\href {https://doi.org/https://doi.org/10.1128/jb.181.20.6419-6424.1999} {\bibfield  {journal} {\bibinfo  {journal} {J. Bacteriol.}\ }\textbf {\bibinfo {volume} {181}},\ \bibinfo {pages} {6419} (\bibinfo {year} {1999})}\BibitemShut {NoStop}%
\bibitem [{\citenamefont {Hu}\ and\ \citenamefont {Lutkenhaus}(1999)}]{hu1999topological}%
  \BibitemOpen
  \bibfield  {author} {\bibinfo {author} {\bibfnamefont {Z.}~\bibnamefont {Hu}}\ and\ \bibinfo {author} {\bibfnamefont {J.}~\bibnamefont {Lutkenhaus}},\ }\bibfield  {title} {\bibinfo {title} {Topological regulation of cell division in escherichia coli involves rapid pole to pole oscillation of the division inhibitor minc under the control of mind and mine},\ }\href {https://doi.org/10.1046/j.1365-2958.1999.01575.x} {\bibfield  {journal} {\bibinfo  {journal} {Mol. Microbiol.}\ }\textbf {\bibinfo {volume} {34}},\ \bibinfo {pages} {82} (\bibinfo {year} {1999})}\BibitemShut {NoStop}%
\bibitem [{\citenamefont {Vecchiarelli}\ \emph {et~al.}(2016)\citenamefont {Vecchiarelli}, \citenamefont {Li}, \citenamefont {Mizuuchi}, \citenamefont {Hwang}, \citenamefont {Seol}, \citenamefont {Neuman},\ and\ \citenamefont {Mizuuchi}}]{vecchiarelli2016membrane}%
  \BibitemOpen
  \bibfield  {author} {\bibinfo {author} {\bibfnamefont {A.~G.}\ \bibnamefont {Vecchiarelli}}, \bibinfo {author} {\bibfnamefont {M.}~\bibnamefont {Li}}, \bibinfo {author} {\bibfnamefont {M.}~\bibnamefont {Mizuuchi}}, \bibinfo {author} {\bibfnamefont {L.~C.}\ \bibnamefont {Hwang}}, \bibinfo {author} {\bibfnamefont {Y.}~\bibnamefont {Seol}}, \bibinfo {author} {\bibfnamefont {K.~C.}\ \bibnamefont {Neuman}},\ and\ \bibinfo {author} {\bibfnamefont {K.}~\bibnamefont {Mizuuchi}},\ }\bibfield  {title} {\bibinfo {title} {Membrane-bound minde complex acts as a toggle switch that drives min oscillation coupled to cytoplasmic depletion of mind},\ }\href {https://doi.org/10.1073/pnas.1600644113} {\bibfield  {journal} {\bibinfo  {journal} {Proc. Nat. Acad. Sci.}\ }\textbf {\bibinfo {volume} {113}},\ \bibinfo {pages} {E1479} (\bibinfo {year} {2016})}\BibitemShut {NoStop}%
\bibitem [{\citenamefont {Ivlev}\ \emph {et~al.}(2015)\citenamefont {Ivlev}, \citenamefont {Bartnick}, \citenamefont {Heinen}, \citenamefont {Du}, \citenamefont {Nosenko},\ and\ \citenamefont {Löwen}}]{Ivlev2015}%
  \BibitemOpen
  \bibfield  {author} {\bibinfo {author} {\bibfnamefont {A.~V.}\ \bibnamefont {Ivlev}}, \bibinfo {author} {\bibfnamefont {J.}~\bibnamefont {Bartnick}}, \bibinfo {author} {\bibfnamefont {M.}~\bibnamefont {Heinen}}, \bibinfo {author} {\bibfnamefont {C.-R.}\ \bibnamefont {Du}}, \bibinfo {author} {\bibfnamefont {V.}~\bibnamefont {Nosenko}},\ and\ \bibinfo {author} {\bibfnamefont {H.}~\bibnamefont {Löwen}},\ }\bibfield  {title} {\bibinfo {title} {Statistical mechanics where newton's third law is broken},\ }\href {https://doi.org/10.1103/PhysRevX.5.011035} {\bibfield  {journal} {\bibinfo  {journal} {Phys. Rev. X}\ }\textbf {\bibinfo {volume} {5}},\ \bibinfo {pages} {11035} (\bibinfo {year} {2015})}\BibitemShut {NoStop}%
\bibitem [{\citenamefont {Agudo-Canalejo}\ and\ \citenamefont {Golestanian}(2019)}]{Agudo2019}%
  \BibitemOpen
  \bibfield  {author} {\bibinfo {author} {\bibfnamefont {J.}~\bibnamefont {Agudo-Canalejo}}\ and\ \bibinfo {author} {\bibfnamefont {R.}~\bibnamefont {Golestanian}},\ }\bibfield  {title} {\bibinfo {title} {Active phase separation in mixtures of chemically interacting particles},\ }\href {https://doi.org/10.1103/PhysRevLett.123.018101} {\bibfield  {journal} {\bibinfo  {journal} {Phys. Rev. Lett.}\ }\textbf {\bibinfo {volume} {123}},\ \bibinfo {pages} {18101} (\bibinfo {year} {2019})}\BibitemShut {NoStop}%
\bibitem [{\citenamefont {Fruchart}\ \emph {et~al.}(2021)\citenamefont {Fruchart}, \citenamefont {Hanai}, \citenamefont {Littlewood},\ and\ \citenamefont {Vitelli}}]{Fruchart2021}%
  \BibitemOpen
  \bibfield  {author} {\bibinfo {author} {\bibfnamefont {M.}~\bibnamefont {Fruchart}}, \bibinfo {author} {\bibfnamefont {R.}~\bibnamefont {Hanai}}, \bibinfo {author} {\bibfnamefont {P.~B.}\ \bibnamefont {Littlewood}},\ and\ \bibinfo {author} {\bibfnamefont {V.}~\bibnamefont {Vitelli}},\ }\bibfield  {title} {\bibinfo {title} {Non-reciprocal phase transitions},\ }\href {https://doi.org/10.1038/s41586-021-03375-9} {\bibfield  {journal} {\bibinfo  {journal} {Nature}\ }\textbf {\bibinfo {volume} {592}},\ \bibinfo {pages} {363} (\bibinfo {year} {2021})}\BibitemShut {NoStop}%
\bibitem [{\citenamefont {Chiu}\ and\ \citenamefont {Omar}(2023)}]{Chiu2023}%
  \BibitemOpen
  \bibfield  {author} {\bibinfo {author} {\bibfnamefont {Y.-J.}\ \bibnamefont {Chiu}}\ and\ \bibinfo {author} {\bibfnamefont {A.~K.}\ \bibnamefont {Omar}},\ }\bibfield  {title} {\bibinfo {title} {Phase coexistence implications of violating newton’s third law},\ }\href {https://doi.org/10.1063/5.0146822} {\bibfield  {journal} {\bibinfo  {journal} {J. Chem. Phys.}\ }\textbf {\bibinfo {volume} {158}} (\bibinfo {year} {2023})}\BibitemShut {NoStop}%
\bibitem [{\citenamefont {Duan}\ \emph {et~al.}(2023)\citenamefont {Duan}, \citenamefont {Agudo-Canalejo}, \citenamefont {Golestanian},\ and\ \citenamefont {Mahault}}]{Duan2023}%
  \BibitemOpen
  \bibfield  {author} {\bibinfo {author} {\bibfnamefont {Y.}~\bibnamefont {Duan}}, \bibinfo {author} {\bibfnamefont {J.}~\bibnamefont {Agudo-Canalejo}}, \bibinfo {author} {\bibfnamefont {R.}~\bibnamefont {Golestanian}},\ and\ \bibinfo {author} {\bibfnamefont {B.}~\bibnamefont {Mahault}},\ }\bibfield  {title} {\bibinfo {title} {Dynamical pattern formation without self-attraction in quorum-sensing active matter: The interplay between nonreciprocity and motility},\ }\href {https://doi.org/10.1103/PhysRevLett.131.148301} {\bibfield  {journal} {\bibinfo  {journal} {Phys. Rev. Lett.}\ }\textbf {\bibinfo {volume} {131}},\ \bibinfo {pages} {148301} (\bibinfo {year} {2023})}\BibitemShut {NoStop}%
\bibitem [{\citenamefont {Duan}\ \emph {et~al.}(2025)\citenamefont {Duan}, \citenamefont {Agudo-Canalejo}, \citenamefont {Golestanian},\ and\ \citenamefont {Mahault}}]{Duan2025}%
  \BibitemOpen
  \bibfield  {author} {\bibinfo {author} {\bibfnamefont {Y.}~\bibnamefont {Duan}}, \bibinfo {author} {\bibfnamefont {J.}~\bibnamefont {Agudo-Canalejo}}, \bibinfo {author} {\bibfnamefont {R.}~\bibnamefont {Golestanian}},\ and\ \bibinfo {author} {\bibfnamefont {B.}~\bibnamefont {Mahault}},\ }\bibfield  {title} {\bibinfo {title} {Phase coexistence in nonreciprocal quorum-sensing active matter},\ }\href {https://doi.org/10.1103/PhysRevResearch.7.013234} {\bibfield  {journal} {\bibinfo  {journal} {Phys. Rev. Res.}\ }\textbf {\bibinfo {volume} {7}},\ \bibinfo {pages} {13234} (\bibinfo {year} {2025})}\BibitemShut {NoStop}%
\bibitem [{\citenamefont {Kreienkamp}\ and\ \citenamefont {Klapp}(2024{\natexlab{a}})}]{Kreienkamp2024}%
  \BibitemOpen
  \bibfield  {author} {\bibinfo {author} {\bibfnamefont {K.~L.}\ \bibnamefont {Kreienkamp}}\ and\ \bibinfo {author} {\bibfnamefont {S.~H.~L.}\ \bibnamefont {Klapp}},\ }\bibfield  {title} {\bibinfo {title} {Dynamical structures in phase-separating nonreciprocal polar active mixtures},\ }\href {https://doi.org/10.1103/PhysRevE.110.064135} {\bibfield  {journal} {\bibinfo  {journal} {Phys. Rev. E}\ }\textbf {\bibinfo {volume} {110}},\ \bibinfo {pages} {64135} (\bibinfo {year} {2024}{\natexlab{a}})}\BibitemShut {NoStop}%
\bibitem [{\citenamefont {Kreienkamp}\ and\ \citenamefont {Klapp}(2024{\natexlab{b}})}]{Kreienkamp20242}%
  \BibitemOpen
  \bibfield  {author} {\bibinfo {author} {\bibfnamefont {K.~L.}\ \bibnamefont {Kreienkamp}}\ and\ \bibinfo {author} {\bibfnamefont {S.~H.~L.}\ \bibnamefont {Klapp}},\ }\bibfield  {title} {\bibinfo {title} {Nonreciprocal alignment induces asymmetric clustering in active mixtures},\ }\href {https://doi.org/10.1103/PhysRevLett.133.258303} {\bibfield  {journal} {\bibinfo  {journal} {Phys. Rev. Lett.}\ }\textbf {\bibinfo {volume} {133}},\ \bibinfo {pages} {258303} (\bibinfo {year} {2024}{\natexlab{b}})}\BibitemShut {NoStop}%
\bibitem [{\citenamefont {Navas}\ and\ \citenamefont {Klapp}(2024)}]{Navas2024}%
  \BibitemOpen
  \bibfield  {author} {\bibinfo {author} {\bibfnamefont {S.~F.}\ \bibnamefont {Navas}}\ and\ \bibinfo {author} {\bibfnamefont {S.~H.~L.}\ \bibnamefont {Klapp}},\ }\bibfield  {title} {\bibinfo {title} {Impact of non-reciprocal interactions on colloidal self-assembly with tunable anisotropy},\ }\href {https://pubs.aip.org/aip/jcp/article-abstract/161/5/054908/3306671/Impact-of-non-reciprocal-interactions-on-colloidal} {\bibfield  {journal} {\bibinfo  {journal} {J. Chem. Phys.}\ }\textbf {\bibinfo {volume} {161}} (\bibinfo {year} {2024})}\BibitemShut {NoStop}%
\bibitem [{\citenamefont {Chen}\ \emph {et~al.}(2024)\citenamefont {Chen}, \citenamefont {Lei}, \citenamefont {Xiang}, \citenamefont {Duan}, \citenamefont {Peng},\ and\ \citenamefont {Zhang}}]{Chen2024}%
  \BibitemOpen
  \bibfield  {author} {\bibinfo {author} {\bibfnamefont {J.}~\bibnamefont {Chen}}, \bibinfo {author} {\bibfnamefont {X.}~\bibnamefont {Lei}}, \bibinfo {author} {\bibfnamefont {Y.}~\bibnamefont {Xiang}}, \bibinfo {author} {\bibfnamefont {M.}~\bibnamefont {Duan}}, \bibinfo {author} {\bibfnamefont {X.}~\bibnamefont {Peng}},\ and\ \bibinfo {author} {\bibfnamefont {H.~P.}\ \bibnamefont {Zhang}},\ }\bibfield  {title} {\bibinfo {title} {Emergent chirality and hyperuniformity in an active mixture with nonreciprocal interactions},\ }\href {https://doi.org/10.1103/PhysRevLett.132.118301} {\bibfield  {journal} {\bibinfo  {journal} {Phys. Rev. Lett.}\ }\textbf {\bibinfo {volume} {132}},\ \bibinfo {pages} {118301} (\bibinfo {year} {2024})}\BibitemShut {NoStop}%
\bibitem [{\citenamefont {Stenhammar}\ \emph {et~al.}(2015)\citenamefont {Stenhammar}, \citenamefont {Wittkowski}, \citenamefont {Marenduzzo},\ and\ \citenamefont {Cates}}]{Stenhammer2015}%
  \BibitemOpen
  \bibfield  {author} {\bibinfo {author} {\bibfnamefont {J.}~\bibnamefont {Stenhammar}}, \bibinfo {author} {\bibfnamefont {R.}~\bibnamefont {Wittkowski}}, \bibinfo {author} {\bibfnamefont {D.}~\bibnamefont {Marenduzzo}},\ and\ \bibinfo {author} {\bibfnamefont {M.~E.}\ \bibnamefont {Cates}},\ }\bibfield  {title} {\bibinfo {title} {Activity-induced phase separation and self-assembly in mixtures of active and passive particles},\ }\href {https://doi.org/10.1103/PhysRevLett.114.018301} {\bibfield  {journal} {\bibinfo  {journal} {Phys. Rev. Lett.}\ }\textbf {\bibinfo {volume} {114}},\ \bibinfo {pages} {18301} (\bibinfo {year} {2015})}\BibitemShut {NoStop}%
\bibitem [{\citenamefont {Wysocki}\ \emph {et~al.}(2016)\citenamefont {Wysocki}, \citenamefont {Winkler},\ and\ \citenamefont {Gompper}}]{Wysocki2016}%
  \BibitemOpen
  \bibfield  {author} {\bibinfo {author} {\bibfnamefont {A.}~\bibnamefont {Wysocki}}, \bibinfo {author} {\bibfnamefont {R.~G.}\ \bibnamefont {Winkler}},\ and\ \bibinfo {author} {\bibfnamefont {G.}~\bibnamefont {Gompper}},\ }\bibfield  {title} {\bibinfo {title} {Propagating interfaces in mixtures of active and passive brownian particles},\ }\href {https://doi.org/10.1088/1367-2630/aa529d} {\bibfield  {journal} {\bibinfo  {journal} {New J. Phys.}\ }\textbf {\bibinfo {volume} {18}},\ \bibinfo {pages} {123030} (\bibinfo {year} {2016})}\BibitemShut {NoStop}%
\bibitem [{\citenamefont {Wittkowski}\ \emph {et~al.}(2017)\citenamefont {Wittkowski}, \citenamefont {Stenhammar},\ and\ \citenamefont {Cates}}]{Wittkowski2017}%
  \BibitemOpen
  \bibfield  {author} {\bibinfo {author} {\bibfnamefont {R.}~\bibnamefont {Wittkowski}}, \bibinfo {author} {\bibfnamefont {J.}~\bibnamefont {Stenhammar}},\ and\ \bibinfo {author} {\bibfnamefont {M.~E.}\ \bibnamefont {Cates}},\ }\bibfield  {title} {\bibinfo {title} {Nonequilibrium dynamics of mixtures of active and passive colloidal particles},\ }\href {https://doi.org/10.1088/1367-2630/aa8195} {\bibfield  {journal} {\bibinfo  {journal} {New J. Phys.}\ }\textbf {\bibinfo {volume} {19}},\ \bibinfo {pages} {105003} (\bibinfo {year} {2017})}\BibitemShut {NoStop}%
\bibitem [{\citenamefont {Agrawal}\ \emph {et~al.}(2017)\citenamefont {Agrawal}, \citenamefont {Bruss},\ and\ \citenamefont {Glotzer}}]{Agrawal2017}%
  \BibitemOpen
  \bibfield  {author} {\bibinfo {author} {\bibfnamefont {M.}~\bibnamefont {Agrawal}}, \bibinfo {author} {\bibfnamefont {I.~R.}\ \bibnamefont {Bruss}},\ and\ \bibinfo {author} {\bibfnamefont {S.~C.}\ \bibnamefont {Glotzer}},\ }\bibfield  {title} {\bibinfo {title} {Tunable emergent structures and traveling waves in mixtures of passive and contact-triggered-active particles},\ }\href {https://doi.org/10.1039/C7SM00888K} {\bibfield  {journal} {\bibinfo  {journal} {Soft Matter}\ }\textbf {\bibinfo {volume} {13}},\ \bibinfo {pages} {6332} (\bibinfo {year} {2017})}\BibitemShut {NoStop}%
\bibitem [{\citenamefont {Banerjee}\ \emph {et~al.}(2022)\citenamefont {Banerjee}, \citenamefont {Mandal}, \citenamefont {Banerjee}, \citenamefont {Thutupalli},\ and\ \citenamefont {Rao}}]{Banjeree2022}%
  \BibitemOpen
  \bibfield  {author} {\bibinfo {author} {\bibfnamefont {J.~P.}\ \bibnamefont {Banerjee}}, \bibinfo {author} {\bibfnamefont {R.}~\bibnamefont {Mandal}}, \bibinfo {author} {\bibfnamefont {D.~S.}\ \bibnamefont {Banerjee}}, \bibinfo {author} {\bibfnamefont {S.}~\bibnamefont {Thutupalli}},\ and\ \bibinfo {author} {\bibfnamefont {M.}~\bibnamefont {Rao}},\ }\bibfield  {title} {\bibinfo {title} {Unjamming and emergent nonreciprocity in active ploughing through a compressible viscoelastic fluid},\ }\href {https://doi.org/10.1038/s41467-022-31984-z} {\bibfield  {journal} {\bibinfo  {journal} {Nat. Comm.}\ }\textbf {\bibinfo {volume} {13}},\ \bibinfo {pages} {4533} (\bibinfo {year} {2022})}\BibitemShut {NoStop}%
\bibitem [{\citenamefont {Dinelli}\ \emph {et~al.}(2023)\citenamefont {Dinelli}, \citenamefont {O’Byrne}, \citenamefont {Curatolo}, \citenamefont {Zhao}, \citenamefont {Sollich},\ and\ \citenamefont {Tailleur}}]{Dinelli2023}%
  \BibitemOpen
  \bibfield  {author} {\bibinfo {author} {\bibfnamefont {A.}~\bibnamefont {Dinelli}}, \bibinfo {author} {\bibfnamefont {J.}~\bibnamefont {O’Byrne}}, \bibinfo {author} {\bibfnamefont {A.}~\bibnamefont {Curatolo}}, \bibinfo {author} {\bibfnamefont {Y.}~\bibnamefont {Zhao}}, \bibinfo {author} {\bibfnamefont {P.}~\bibnamefont {Sollich}},\ and\ \bibinfo {author} {\bibfnamefont {J.}~\bibnamefont {Tailleur}},\ }\bibfield  {title} {\bibinfo {title} {Non-reciprocity across scales in active mixtures},\ }\href {https://doi.org/10.1038/s41467-023-42713-5} {\bibfield  {journal} {\bibinfo  {journal} {Nat. Comm.}\ }\textbf {\bibinfo {volume} {14}},\ \bibinfo {pages} {7035} (\bibinfo {year} {2023})}\BibitemShut {NoStop}%
\bibitem [{\citenamefont {Tucci}\ \emph {et~al.}(2024)\citenamefont {Tucci}, \citenamefont {Golestanian},\ and\ \citenamefont {Saha}}]{Tucci2024}%
  \BibitemOpen
  \bibfield  {author} {\bibinfo {author} {\bibfnamefont {G.}~\bibnamefont {Tucci}}, \bibinfo {author} {\bibfnamefont {R.}~\bibnamefont {Golestanian}},\ and\ \bibinfo {author} {\bibfnamefont {S.}~\bibnamefont {Saha}},\ }\bibfield  {title} {\bibinfo {title} {Nonreciprocal collective dynamics in a mixture of phoretic janus colloids},\ }\href {https://doi.org/10.1088/1367-2630/ad50ff} {\bibfield  {journal} {\bibinfo  {journal} {New J. Phys.}\ }\textbf {\bibinfo {volume} {26}},\ \bibinfo {pages} {073006} (\bibinfo {year} {2024})}\BibitemShut {NoStop}%
\bibitem [{\citenamefont {Mason}\ \emph {et~al.}(2025)\citenamefont {Mason}, \citenamefont {Jack},\ and\ \citenamefont {Bruna}}]{Mason2024}%
  \BibitemOpen
  \bibfield  {author} {\bibinfo {author} {\bibfnamefont {J.}~\bibnamefont {Mason}}, \bibinfo {author} {\bibfnamefont {R.~L.}\ \bibnamefont {Jack}},\ and\ \bibinfo {author} {\bibfnamefont {M.}~\bibnamefont {Bruna}},\ }\bibfield  {title} {\bibinfo {title} {Dynamical patterns and nonreciprocal effective interactions in an active-passive mixture through exact hydrodynamic analysis},\ }\href {https://doi.org/10.1038/s41467-025-60518-6} {\bibfield  {journal} {\bibinfo  {journal} {Nat. Comm.}\ }\textbf {\bibinfo {volume} {16}},\ \bibinfo {pages} {6017} (\bibinfo {year} {2025})}\BibitemShut {NoStop}%
\bibitem [{\citenamefont {Saha}\ \emph {et~al.}(2020)\citenamefont {Saha}, \citenamefont {Agudo-Canalejo},\ and\ \citenamefont {Golestanian}}]{Saha2020}%
  \BibitemOpen
  \bibfield  {author} {\bibinfo {author} {\bibfnamefont {S.}~\bibnamefont {Saha}}, \bibinfo {author} {\bibfnamefont {J.}~\bibnamefont {Agudo-Canalejo}},\ and\ \bibinfo {author} {\bibfnamefont {R.}~\bibnamefont {Golestanian}},\ }\bibfield  {title} {\bibinfo {title} {Scalar active mixtures: The nonreciprocal cahn-hilliard model},\ }\href {https://doi.org/10.1103/PhysRevX.10.041009} {\bibfield  {journal} {\bibinfo  {journal} {Phys. Rev. X}\ }\textbf {\bibinfo {volume} {10}},\ \bibinfo {pages} {041009} (\bibinfo {year} {2020})}\BibitemShut {NoStop}%
\bibitem [{\citenamefont {You}\ \emph {et~al.}(2020)\citenamefont {You}, \citenamefont {Baskaran},\ and\ \citenamefont {Marchetti}}]{You2020}%
  \BibitemOpen
  \bibfield  {author} {\bibinfo {author} {\bibfnamefont {Z.}~\bibnamefont {You}}, \bibinfo {author} {\bibfnamefont {A.}~\bibnamefont {Baskaran}},\ and\ \bibinfo {author} {\bibfnamefont {M.~C.}\ \bibnamefont {Marchetti}},\ }\bibfield  {title} {\bibinfo {title} {Nonreciprocity as a generic route to traveling states},\ }\href {https://doi.org/10.1073/pnas.2010318117} {\bibfield  {journal} {\bibinfo  {journal} {Proc. Nat. Acad. Sci.}\ }\textbf {\bibinfo {volume} {117}},\ \bibinfo {pages} {19767} (\bibinfo {year} {2020})}\BibitemShut {NoStop}%
\bibitem [{\citenamefont {Brauns}\ and\ \citenamefont {Marchetti}(2024)}]{Brauns2024}%
  \BibitemOpen
  \bibfield  {author} {\bibinfo {author} {\bibfnamefont {F.}~\bibnamefont {Brauns}}\ and\ \bibinfo {author} {\bibfnamefont {M.~C.}\ \bibnamefont {Marchetti}},\ }\bibfield  {title} {\bibinfo {title} {Nonreciprocal pattern formation of conserved fields},\ }\href {https://doi.org/10.1103/PhysRevX.14.021014} {\bibfield  {journal} {\bibinfo  {journal} {Phys. Rev. X}\ }\textbf {\bibinfo {volume} {14}},\ \bibinfo {pages} {021014} (\bibinfo {year} {2024})}\BibitemShut {NoStop}%
\bibitem [{Note1()}]{Note1}%
  \BibitemOpen
  \bibinfo {note} {When considering the dynamics of fields which can take complex values, non-reciprocity is often believed to correspond to a non-Hermitian Jacobian. This again is not a sufficient definition of nonreciprocity for the same reason that the the asymmetry of a real Jacobian is not a unique indicator of nonreciprocity.}\BibitemShut {Stop}%
\bibitem [{\citenamefont {Fruchart}\ and\ \citenamefont {Vitelli}(2026)}]{fruchart2026}%
  \BibitemOpen
  \bibfield  {author} {\bibinfo {author} {\bibfnamefont {M.}~\bibnamefont {Fruchart}}\ and\ \bibinfo {author} {\bibfnamefont {V.}~\bibnamefont {Vitelli}},\ }\bibfield  {title} {\bibinfo {title} {Nonreciprocal many-body physics},\ }\href {https://arxiv.org/abs/2602.11111} {\bibfield  {journal} {\bibinfo  {journal} {arXiv preprint arXiv:2602.11111}\ } (\bibinfo {year} {2026})}\BibitemShut {NoStop}%
\bibitem [{\citenamefont {Horn}\ and\ \citenamefont {Jackson}(1972)}]{Horn1972}%
  \BibitemOpen
  \bibfield  {author} {\bibinfo {author} {\bibfnamefont {F.}~\bibnamefont {Horn}}\ and\ \bibinfo {author} {\bibfnamefont {R.}~\bibnamefont {Jackson}},\ }\bibfield  {title} {\bibinfo {title} {General mass action kinetics},\ }\href {https://doi.org/10.1007/BF00251225} {\bibfield  {journal} {\bibinfo  {journal} {Arch. Rat. Mech. Anal.}\ }\textbf {\bibinfo {volume} {47}},\ \bibinfo {pages} {81} (\bibinfo {year} {1972})}\BibitemShut {NoStop}%
\bibitem [{\citenamefont {Feinberg}(1972)}]{Feinberg1972}%
  \BibitemOpen
  \bibfield  {author} {\bibinfo {author} {\bibfnamefont {M.}~\bibnamefont {Feinberg}},\ }\bibfield  {title} {\bibinfo {title} {Complex balancing in general kinetic systems},\ }\href {https://doi.org/10.1007/BF00255665} {\bibfield  {journal} {\bibinfo  {journal} {Arch. Rat. Mech. Anal.}\ }\textbf {\bibinfo {volume} {49}},\ \bibinfo {pages} {187} (\bibinfo {year} {1972})}\BibitemShut {NoStop}%
\bibitem [{\citenamefont {Avanzini}\ \emph {et~al.}(2021)\citenamefont {Avanzini}, \citenamefont {Penocchio}, \citenamefont {Falasco},\ and\ \citenamefont {Esposito}}]{Avanzini2021}%
  \BibitemOpen
  \bibfield  {author} {\bibinfo {author} {\bibfnamefont {F.}~\bibnamefont {Avanzini}}, \bibinfo {author} {\bibfnamefont {E.}~\bibnamefont {Penocchio}}, \bibinfo {author} {\bibfnamefont {G.}~\bibnamefont {Falasco}},\ and\ \bibinfo {author} {\bibfnamefont {M.}~\bibnamefont {Esposito}},\ }\bibfield  {title} {\bibinfo {title} {Nonequilibrium thermodynamics of non-ideal chemical reaction networks},\ }\href {https://doi.org/10.1063/5.0041225} {\bibfield  {journal} {\bibinfo  {journal} {J. Chem. Phys.}\ }\textbf {\bibinfo {volume} {154}} (\bibinfo {year} {2021})}\BibitemShut {NoStop}%
\bibitem [{\citenamefont {Avanzini}\ \emph {et~al.}(2024)\citenamefont {Avanzini}, \citenamefont {Aslyamov}, \citenamefont {Étienne Fodor},\ and\ \citenamefont {Esposito}}]{Avanzini2024}%
  \BibitemOpen
  \bibfield  {author} {\bibinfo {author} {\bibfnamefont {F.}~\bibnamefont {Avanzini}}, \bibinfo {author} {\bibfnamefont {T.}~\bibnamefont {Aslyamov}}, \bibinfo {author} {\bibnamefont {Étienne Fodor}},\ and\ \bibinfo {author} {\bibfnamefont {M.}~\bibnamefont {Esposito}},\ }\bibfield  {title} {\bibinfo {title} {Nonequilibrium thermodynamics of non-ideal reaction–diffusion systems: Implications for active self-organization},\ }\href {https://doi.org/10.1063/5.0231520} {\bibfield  {journal} {\bibinfo  {journal} {J. Chem. Phys.}\ }\textbf {\bibinfo {volume} {161}} (\bibinfo {year} {2024})}\BibitemShut {NoStop}%
\bibitem [{\citenamefont {Remlein}\ \emph {et~al.}(2025)\citenamefont {Remlein}, \citenamefont {Esposito},\ and\ \citenamefont {Avanzini}}]{Remlein2025}%
  \BibitemOpen
  \bibfield  {author} {\bibinfo {author} {\bibfnamefont {B.}~\bibnamefont {Remlein}}, \bibinfo {author} {\bibfnamefont {M.}~\bibnamefont {Esposito}},\ and\ \bibinfo {author} {\bibfnamefont {F.}~\bibnamefont {Avanzini}},\ }\bibfield  {title} {\bibinfo {title} {What is a chemostat? insights from hybrid dynamics and stochastic thermodynamics},\ }\href {https://pubs.aip.org/aip/jcp/article-abstract/162/22/224113/3349837/What-is-a-chemostat-Insights-from-hybrid-dynamics?redirectedFrom=fulltext} {\bibfield  {journal} {\bibinfo  {journal} {J. Chem. Phys.}\ }\textbf {\bibinfo {volume} {162}} (\bibinfo {year} {2025})}\BibitemShut {NoStop}%
\bibitem [{Note2()}]{Note2}%
  \BibitemOpen
  \bibinfo {note} {See the Supplemental Material (SM) for supporting examples and derivations, which includes discussions related to Refs.~\cite {murugan2017topologically, Tang2021, Ezawa2022, Nelson2024, zheng2024topological, veenstra2024non, Belyansky2025}}\BibitemShut {NoStop}%
\bibitem [{\citenamefont {Polettini}\ and\ \citenamefont {Esposito}(2014)}]{Polettini2014}%
  \BibitemOpen
  \bibfield  {author} {\bibinfo {author} {\bibfnamefont {M.}~\bibnamefont {Polettini}}\ and\ \bibinfo {author} {\bibfnamefont {M.}~\bibnamefont {Esposito}},\ }\bibfield  {title} {\bibinfo {title} {Irreversible thermodynamics of open chemical networks. i. emergent cycles and broken conservation laws},\ }\href {https://pubs.aip.org/aip/jcp/article-abstract/141/2/024117/352438/Irreversible-thermodynamics-of-open-chemical} {\bibfield  {journal} {\bibinfo  {journal} {J. Chem. Phys.}\ }\textbf {\bibinfo {volume} {141}} (\bibinfo {year} {2014})}\BibitemShut {NoStop}%
\bibitem [{\citenamefont {Rao}\ and\ \citenamefont {Esposito}(2016)}]{Rao2016}%
  \BibitemOpen
  \bibfield  {author} {\bibinfo {author} {\bibfnamefont {R.}~\bibnamefont {Rao}}\ and\ \bibinfo {author} {\bibfnamefont {M.}~\bibnamefont {Esposito}},\ }\bibfield  {title} {\bibinfo {title} {Nonequilibrium thermodynamics of chemical reaction networks: Wisdom from stochastic thermodynamics},\ }\href {https://doi.org/10.1103/PhysRevX.6.041064} {\bibfield  {journal} {\bibinfo  {journal} {Phys. Rev. X}\ }\textbf {\bibinfo {volume} {6}},\ \bibinfo {pages} {041064} (\bibinfo {year} {2016})}\BibitemShut {NoStop}%
\bibitem [{\citenamefont {Groot}\ and\ \citenamefont {Mazur}(2013)}]{degroot2013}%
  \BibitemOpen
  \bibfield  {author} {\bibinfo {author} {\bibfnamefont {S.~R.~D.}\ \bibnamefont {Groot}}\ and\ \bibinfo {author} {\bibfnamefont {P.}~\bibnamefont {Mazur}},\ }\href@noop {} {\emph {\bibinfo {title} {Non-equilibrium thermodynamics}}}\ (\bibinfo  {publisher} {Courier Corporation},\ \bibinfo {year} {2013})\BibitemShut {NoStop}%
\bibitem [{\citenamefont {Onsager}(1931{\natexlab{a}})}]{Onsager1931A}%
  \BibitemOpen
  \bibfield  {author} {\bibinfo {author} {\bibfnamefont {L.}~\bibnamefont {Onsager}},\ }\bibfield  {title} {\bibinfo {title} {Reciprocal relations in irreversible processes. i.},\ }\href {https://doi.org/10.1103/PhysRev.37.405} {\bibfield  {journal} {\bibinfo  {journal} {Phys. Rev.}\ }\textbf {\bibinfo {volume} {37}},\ \bibinfo {pages} {405} (\bibinfo {year} {1931}{\natexlab{a}})}\BibitemShut {NoStop}%
\bibitem [{\citenamefont {Onsager}(1931{\natexlab{b}})}]{Onsager1931B}%
  \BibitemOpen
  \bibfield  {author} {\bibinfo {author} {\bibfnamefont {L.}~\bibnamefont {Onsager}},\ }\bibfield  {title} {\bibinfo {title} {Reciprocal relations in irreversible processes. ii.},\ }\href {https://doi.org/10.1103/PhysRev.38.2265} {\bibfield  {journal} {\bibinfo  {journal} {Phys. Rev.}\ }\textbf {\bibinfo {volume} {38}},\ \bibinfo {pages} {2265} (\bibinfo {year} {1931}{\natexlab{b}})}\BibitemShut {NoStop}%
\bibitem [{\citenamefont {Bazant}(2013)}]{Bazant2013}%
  \BibitemOpen
  \bibfield  {author} {\bibinfo {author} {\bibfnamefont {M.~Z.}\ \bibnamefont {Bazant}},\ }\bibfield  {title} {\bibinfo {title} {Theory of chemical kinetics and charge transfer based on nonequilibrium thermodynamics},\ }\href {https://doi.org/10.1021/ar300145c} {\bibfield  {journal} {\bibinfo  {journal} {Acc. Chem. Res.}\ }\textbf {\bibinfo {volume} {46}},\ \bibinfo {pages} {1144} (\bibinfo {year} {2013})}\BibitemShut {NoStop}%
\bibitem [{\citenamefont {Weber}\ \emph {et~al.}(2019)\citenamefont {Weber}, \citenamefont {Zwicker}, \citenamefont {J{\"u}licher},\ and\ \citenamefont {Lee}}]{weber2019physics}%
  \BibitemOpen
  \bibfield  {author} {\bibinfo {author} {\bibfnamefont {C.~A.}\ \bibnamefont {Weber}}, \bibinfo {author} {\bibfnamefont {D.}~\bibnamefont {Zwicker}}, \bibinfo {author} {\bibfnamefont {F.}~\bibnamefont {J{\"u}licher}},\ and\ \bibinfo {author} {\bibfnamefont {C.~F.}\ \bibnamefont {Lee}},\ }\bibfield  {title} {\bibinfo {title} {Physics of active emulsions},\ }\href {https://doi.org/10.1088/1361-6633/ab052b} {\bibfield  {journal} {\bibinfo  {journal} {Reports on Progress in Physics}\ }\textbf {\bibinfo {volume} {82}},\ \bibinfo {pages} {064601} (\bibinfo {year} {2019})}\BibitemShut {NoStop}%
\bibitem [{\citenamefont {Kirschbaum}\ and\ \citenamefont {Zwicker}(2021)}]{Kirschbaum2021}%
  \BibitemOpen
  \bibfield  {author} {\bibinfo {author} {\bibfnamefont {J.}~\bibnamefont {Kirschbaum}}\ and\ \bibinfo {author} {\bibfnamefont {D.}~\bibnamefont {Zwicker}},\ }\bibfield  {title} {\bibinfo {title} {Controlling biomolecular condensates via chemical reactions},\ }\href {https://doi.org/10.1098/rsif.2021.0255} {\bibfield  {journal} {\bibinfo  {journal} {J. R. Soc. Interface}\ }\textbf {\bibinfo {volume} {18}},\ \bibinfo {pages} {20210255} (\bibinfo {year} {2021})}\BibitemShut {NoStop}%
\bibitem [{\citenamefont {Bauermann}\ \emph {et~al.}(2022)\citenamefont {Bauermann}, \citenamefont {Weber},\ and\ \citenamefont {J{\"u}licher}}]{bauermann2022energy}%
  \BibitemOpen
  \bibfield  {author} {\bibinfo {author} {\bibfnamefont {J.}~\bibnamefont {Bauermann}}, \bibinfo {author} {\bibfnamefont {C.~A.}\ \bibnamefont {Weber}},\ and\ \bibinfo {author} {\bibfnamefont {F.}~\bibnamefont {J{\"u}licher}},\ }\bibfield  {title} {\bibinfo {title} {Energy and matter supply for active droplets},\ }\href {https://doi.org/10.1002/andp.202200132} {\bibfield  {journal} {\bibinfo  {journal} {Annalen der Physik}\ }\textbf {\bibinfo {volume} {534}},\ \bibinfo {pages} {2200132} (\bibinfo {year} {2022})}\BibitemShut {NoStop}%
\bibitem [{\citenamefont {Aslyamov}\ \emph {et~al.}(2023)\citenamefont {Aslyamov}, \citenamefont {Avanzini}, \citenamefont {Étienne Fodor},\ and\ \citenamefont {Esposito}}]{Aslyamov2023}%
  \BibitemOpen
  \bibfield  {author} {\bibinfo {author} {\bibfnamefont {T.}~\bibnamefont {Aslyamov}}, \bibinfo {author} {\bibfnamefont {F.}~\bibnamefont {Avanzini}}, \bibinfo {author} {\bibnamefont {Étienne Fodor}},\ and\ \bibinfo {author} {\bibfnamefont {M.}~\bibnamefont {Esposito}},\ }\bibfield  {title} {\bibinfo {title} {Nonideal reaction-diffusion systems: Multiple routes to instability},\ }\href {https://doi.org/10.1103/PhysRevLett.131.138301} {\bibfield  {journal} {\bibinfo  {journal} {Phys. Rev. Lett.}\ }\textbf {\bibinfo {volume} {131}},\ \bibinfo {pages} {138301} (\bibinfo {year} {2023})}\BibitemShut {NoStop}%
\bibitem [{\citenamefont {Donder}(1927)}]{DeDonder1927}%
  \BibitemOpen
  \bibfield  {author} {\bibinfo {author} {\bibfnamefont {T.~D.}\ \bibnamefont {Donder}},\ }\href@noop {} {\emph {\bibinfo {title} {L'affinité}}},\ Vol.~\bibinfo {volume} {1}\ (\bibinfo  {publisher} {Gauthier-Villars},\ \bibinfo {year} {1927})\BibitemShut {NoStop}%
\bibitem [{Note3()}]{Note3}%
  \BibitemOpen
  \bibinfo {note} {This corresponds to assuming $n$-body reactions are purely $n$-body events that are unaffected by other degrees of freedom. If $k_r$ were to depend on $\protect \bm {\rho }$, this dependence would not enter the linearized dynamics near DB steady states but would be present near CB and non-CB steady states.}\BibitemShut {Stop}%
\bibitem [{Note4()}]{Note4}%
  \BibitemOpen
  \bibinfo {note} {If this series is truncated at order $n$, $\protect \mathbf {H}_n$ must be PD to ensure infinitesimally small fluctuations (i.e.,~$\delta \protect \tilde {\protect \bm {\rho }}_{q}$ in the limit $q \rightarrow \infty $) are penalized.}\BibitemShut {Stop}%
\bibitem [{\citenamefont {Villani}(2009)}]{villani2009hypocoercivity}%
  \BibitemOpen
  \bibfield  {author} {\bibinfo {author} {\bibfnamefont {C.}~\bibnamefont {Villani}},\ }\href {https://doi.org/10.1090/S0065-9266-09-00567-5} {\emph {\bibinfo {title} {Hypocoercivity}}},\ \bibinfo {series} {Hypocoercivity}\ No.\ \bibinfo {number} {nos. 949-951}\ (\bibinfo  {publisher} {American Mathematical Society},\ \bibinfo {year} {2009})\BibitemShut {NoStop}%
\bibitem [{Note5()}]{Note5}%
  \BibitemOpen
  \bibinfo {note} {This was generalized to non-ideal chemostats in Ref.~\cite {Avanzini2024} by introducing auxilliary, ideal chemostats that the the non-ideal chemostats equilibrate with on fast time scales relative to other reactions.}\BibitemShut {Stop}%
\bibitem [{\citenamefont {Hatano}\ and\ \citenamefont {Nelson}(1996)}]{Hatano1996}%
  \BibitemOpen
  \bibfield  {author} {\bibinfo {author} {\bibfnamefont {N.}~\bibnamefont {Hatano}}\ and\ \bibinfo {author} {\bibfnamefont {D.~R.}\ \bibnamefont {Nelson}},\ }\bibfield  {title} {\bibinfo {title} {Localization transitions in non-hermitian quantum mechanics},\ }\href {https://doi.org/10.1103/PhysRevLett.77.570} {\bibfield  {journal} {\bibinfo  {journal} {Phys. Rev. Lett.}\ }\textbf {\bibinfo {volume} {77}},\ \bibinfo {pages} {570} (\bibinfo {year} {1996})}\BibitemShut {NoStop}%
\bibitem [{\citenamefont {Hatano}\ and\ \citenamefont {Nelson}(1997)}]{Hatano1997}%
  \BibitemOpen
  \bibfield  {author} {\bibinfo {author} {\bibfnamefont {N.}~\bibnamefont {Hatano}}\ and\ \bibinfo {author} {\bibfnamefont {D.~R.}\ \bibnamefont {Nelson}},\ }\bibfield  {title} {\bibinfo {title} {Vortex pinning and non-hermitian quantum mechanics},\ }\href {https://doi.org/10.1103/PhysRevB.56.8651} {\bibfield  {journal} {\bibinfo  {journal} {Phys. Rev. B}\ }\textbf {\bibinfo {volume} {56}},\ \bibinfo {pages} {8651} (\bibinfo {year} {1997})}\BibitemShut {NoStop}%
\bibitem [{\citenamefont {Zhang}\ \emph {et~al.}(2022)\citenamefont {Zhang}, \citenamefont {Zhang}, \citenamefont {Lu},\ and\ \citenamefont {Chen}}]{Zhang2022}%
  \BibitemOpen
  \bibfield  {author} {\bibinfo {author} {\bibfnamefont {X.}~\bibnamefont {Zhang}}, \bibinfo {author} {\bibfnamefont {T.}~\bibnamefont {Zhang}}, \bibinfo {author} {\bibfnamefont {M.-H.}\ \bibnamefont {Lu}},\ and\ \bibinfo {author} {\bibfnamefont {Y.-F.}\ \bibnamefont {Chen}},\ }\bibfield  {title} {\bibinfo {title} {A review on non-hermitian skin effect},\ }\href {https://doi.org/10.1080/23746149.2022.2109431} {\bibfield  {journal} {\bibinfo  {journal} {Advances in Physics: X}\ }\textbf {\bibinfo {volume} {7}},\ \bibinfo {pages} {2109431} (\bibinfo {year} {2022})}\BibitemShut {NoStop}%
\bibitem [{\citenamefont {Yokomizo}\ and\ \citenamefont {Murakami}(2019)}]{Yokomizo2019}%
  \BibitemOpen
  \bibfield  {author} {\bibinfo {author} {\bibfnamefont {K.}~\bibnamefont {Yokomizo}}\ and\ \bibinfo {author} {\bibfnamefont {S.}~\bibnamefont {Murakami}},\ }\bibfield  {title} {\bibinfo {title} {Non-bloch band theory of non-hermitian systems},\ }\href {https://doi.org/10.1103/PhysRevLett.123.066404} {\bibfield  {journal} {\bibinfo  {journal} {Phys. Rev. Lett.}\ }\textbf {\bibinfo {volume} {123}},\ \bibinfo {pages} {066404} (\bibinfo {year} {2019})}\BibitemShut {NoStop}%
\bibitem [{\citenamefont {Zhang}\ \emph {et~al.}(2020)\citenamefont {Zhang}, \citenamefont {Yang},\ and\ \citenamefont {Fang}}]{Zhang2020}%
  \BibitemOpen
  \bibfield  {author} {\bibinfo {author} {\bibfnamefont {K.}~\bibnamefont {Zhang}}, \bibinfo {author} {\bibfnamefont {Z.}~\bibnamefont {Yang}},\ and\ \bibinfo {author} {\bibfnamefont {C.}~\bibnamefont {Fang}},\ }\bibfield  {title} {\bibinfo {title} {Correspondence between winding numbers and skin modes in non-hermitian systems},\ }\href {https://doi.org/10.1103/PhysRevLett.125.126402} {\bibfield  {journal} {\bibinfo  {journal} {Phys. Rev. Lett.}\ }\textbf {\bibinfo {volume} {125}},\ \bibinfo {pages} {126402} (\bibinfo {year} {2020})}\BibitemShut {NoStop}%
\bibitem [{\citenamefont {Poggioli}\ and\ \citenamefont {Limmer}(2023)}]{poggioli2023}%
  \BibitemOpen
  \bibfield  {author} {\bibinfo {author} {\bibfnamefont {A.~R.}\ \bibnamefont {Poggioli}}\ and\ \bibinfo {author} {\bibfnamefont {D.~T.}\ \bibnamefont {Limmer}},\ }\bibfield  {title} {\bibinfo {title} {Odd mobility of a passive tracer in a chiral active fluid},\ }\href {https://doi.org/10.1103/PhysRevLett.130.158201} {\bibfield  {journal} {\bibinfo  {journal} {Phys. Rev. Lett.}\ }\textbf {\bibinfo {volume} {130}},\ \bibinfo {pages} {158201} (\bibinfo {year} {2023})}\BibitemShut {NoStop}%
\bibitem [{\citenamefont {Hwang}\ \emph {et~al.}(1993)\citenamefont {Hwang}, \citenamefont {Hwang-Ma},\ and\ \citenamefont {Sheu}}]{hwang1993accelerating}%
  \BibitemOpen
  \bibfield  {author} {\bibinfo {author} {\bibfnamefont {C.-R.}\ \bibnamefont {Hwang}}, \bibinfo {author} {\bibfnamefont {S.-Y.}\ \bibnamefont {Hwang-Ma}},\ and\ \bibinfo {author} {\bibfnamefont {S.-J.}\ \bibnamefont {Sheu}},\ }\bibfield  {title} {\bibinfo {title} {Accelerating gaussian diffusions},\ }\href {https://doi.org/10.1214/aoap/1177005371} {\bibfield  {journal} {\bibinfo  {journal} {Ann. Appl. Probab.}\ ,\ \bibinfo {pages} {897}} (\bibinfo {year} {1993})}\BibitemShut {NoStop}%
\bibitem [{\citenamefont {Hwang}\ \emph {et~al.}(2005)\citenamefont {Hwang}, \citenamefont {Hwang-Ma},\ and\ \citenamefont {Sheu}}]{hwang2005accelerating}%
  \BibitemOpen
  \bibfield  {author} {\bibinfo {author} {\bibfnamefont {C.-R.}\ \bibnamefont {Hwang}}, \bibinfo {author} {\bibfnamefont {S.-Y.}\ \bibnamefont {Hwang-Ma}},\ and\ \bibinfo {author} {\bibfnamefont {S.-J.}\ \bibnamefont {Sheu}},\ }\bibfield  {title} {\bibinfo {title} {Accelerating diffusions},\ }\href {https://arxiv.org/abs/math/0505245v1} {\bibfield  {journal} {\bibinfo  {journal} {arXiv preprint arXiv:math/0505245}\ } (\bibinfo {year} {2005})}\BibitemShut {NoStop}%
\bibitem [{\citenamefont {Ohzeki}\ and\ \citenamefont {Ichiki}(2015)}]{Ohzeki2015}%
  \BibitemOpen
  \bibfield  {author} {\bibinfo {author} {\bibfnamefont {M.}~\bibnamefont {Ohzeki}}\ and\ \bibinfo {author} {\bibfnamefont {A.}~\bibnamefont {Ichiki}},\ }\bibfield  {title} {\bibinfo {title} {Langevin dynamics neglecting detailed balance condition},\ }\href {https://doi.org/10.1103/PhysRevE.92.012105} {\bibfield  {journal} {\bibinfo  {journal} {Phys. Rev. E}\ }\textbf {\bibinfo {volume} {92}},\ \bibinfo {pages} {012105} (\bibinfo {year} {2015})}\BibitemShut {NoStop}%
\bibitem [{\citenamefont {Kaiser}\ \emph {et~al.}(2017)\citenamefont {Kaiser}, \citenamefont {Jack},\ and\ \citenamefont {Zimmer}}]{kaiser2017acceleration}%
  \BibitemOpen
  \bibfield  {author} {\bibinfo {author} {\bibfnamefont {M.}~\bibnamefont {Kaiser}}, \bibinfo {author} {\bibfnamefont {R.~L.}\ \bibnamefont {Jack}},\ and\ \bibinfo {author} {\bibfnamefont {J.}~\bibnamefont {Zimmer}},\ }\bibfield  {title} {\bibinfo {title} {Acceleration of convergence to equilibrium in markov chains by breaking detailed balance},\ }\href {https://doi.org/10.1007/s10955-017-1805-z} {\bibfield  {journal} {\bibinfo  {journal} {J. Stat. Phys.}\ }\textbf {\bibinfo {volume} {168}},\ \bibinfo {pages} {259} (\bibinfo {year} {2017})}\BibitemShut {NoStop}%
\bibitem [{\citenamefont {Gao}\ \emph {et~al.}(2020)\citenamefont {Gao}, \citenamefont {G\"{u}rb\"{u}zbalaban},\ and\ \citenamefont {Zhu}}]{Gao2020}%
  \BibitemOpen
  \bibfield  {author} {\bibinfo {author} {\bibfnamefont {X.}~\bibnamefont {Gao}}, \bibinfo {author} {\bibfnamefont {M.}~\bibnamefont {G\"{u}rb\"{u}zbalaban}},\ and\ \bibinfo {author} {\bibfnamefont {L.}~\bibnamefont {Zhu}},\ }\bibfield  {title} {\bibinfo {title} {Breaking reversibility accelerates langevin dynamics for non-convex optimization},\ }in\ \href {https://dl.acm.org/doi/abs/10.5555/3495724.3497222} {\emph {\bibinfo {booktitle} {Proceedings of the 34th International Conference on Neural Information Processing Systems}}},\ \bibinfo {series and number} {NIPS '20}\ (\bibinfo  {publisher} {Curran Associates Inc.},\ \bibinfo {address} {Red Hook, NY, USA},\ \bibinfo {year} {2020})\BibitemShut {NoStop}%
\bibitem [{\citenamefont {Futami}\ \emph {et~al.}(2020)\citenamefont {Futami}, \citenamefont {Sato},\ and\ \citenamefont {Sugiyama}}]{futami2020}%
  \BibitemOpen
  \bibfield  {author} {\bibinfo {author} {\bibfnamefont {F.}~\bibnamefont {Futami}}, \bibinfo {author} {\bibfnamefont {I.}~\bibnamefont {Sato}},\ and\ \bibinfo {author} {\bibfnamefont {M.}~\bibnamefont {Sugiyama}},\ }\bibfield  {title} {\bibinfo {title} {Accelerating the diffusion-based ensemble sampling by non-reversible dynamics},\ }in\ \href {https://proceedings.mlr.press/v119/futami20a.html} {\emph {\bibinfo {booktitle} {Proceedings of the 37th International Conference on Machine Learning}}},\ \bibinfo {series} {Proc. Mach. Learn. Res.}, Vol.\ \bibinfo {volume} {119},\ \bibinfo {editor} {edited by\ \bibinfo {editor} {\bibfnamefont {H.~D.}\ \bibnamefont {III}}\ and\ \bibinfo {editor} {\bibfnamefont {A.}~\bibnamefont {Singh}}}\ (\bibinfo  {publisher} {PMLR},\ \bibinfo {year} {2020})\ pp.\ \bibinfo {pages} {3337--3347}\BibitemShut {NoStop}%
\bibitem [{\citenamefont {Ghimenti}\ and\ \citenamefont {van Wijland}(2022)}]{Ghimenti2022}%
  \BibitemOpen
  \bibfield  {author} {\bibinfo {author} {\bibfnamefont {F.}~\bibnamefont {Ghimenti}}\ and\ \bibinfo {author} {\bibfnamefont {F.}~\bibnamefont {van Wijland}},\ }\bibfield  {title} {\bibinfo {title} {Accelerating, to some extent, the $p$-spin dynamics},\ }\href {https://doi.org/10.1103/PhysRevE.105.054137} {\bibfield  {journal} {\bibinfo  {journal} {Phys. Rev. E}\ }\textbf {\bibinfo {volume} {105}},\ \bibinfo {pages} {054137} (\bibinfo {year} {2022})}\BibitemShut {NoStop}%
\bibitem [{\citenamefont {Ghimenti}\ \emph {et~al.}(2023)\citenamefont {Ghimenti}, \citenamefont {Berthier}, \citenamefont {Szamel},\ and\ \citenamefont {van Wijland}}]{Ghimenti2023}%
  \BibitemOpen
  \bibfield  {author} {\bibinfo {author} {\bibfnamefont {F.}~\bibnamefont {Ghimenti}}, \bibinfo {author} {\bibfnamefont {L.}~\bibnamefont {Berthier}}, \bibinfo {author} {\bibfnamefont {G.}~\bibnamefont {Szamel}},\ and\ \bibinfo {author} {\bibfnamefont {F.}~\bibnamefont {van Wijland}},\ }\bibfield  {title} {\bibinfo {title} {Sampling efficiency of transverse forces in dense liquids},\ }\href {https://doi.org/10.1103/PhysRevLett.131.257101} {\bibfield  {journal} {\bibinfo  {journal} {Phys. Rev. Lett.}\ }\textbf {\bibinfo {volume} {131}},\ \bibinfo {pages} {257101} (\bibinfo {year} {2023})}\BibitemShut {NoStop}%
\bibitem [{\citenamefont {Ghimenti}\ \emph {et~al.}(2024{\natexlab{a}})\citenamefont {Ghimenti}, \citenamefont {Berthier}, \citenamefont {Szamel},\ and\ \citenamefont {van Wijland}}]{Ghimenti2024a}%
  \BibitemOpen
  \bibfield  {author} {\bibinfo {author} {\bibfnamefont {F.}~\bibnamefont {Ghimenti}}, \bibinfo {author} {\bibfnamefont {L.}~\bibnamefont {Berthier}}, \bibinfo {author} {\bibfnamefont {G.}~\bibnamefont {Szamel}},\ and\ \bibinfo {author} {\bibfnamefont {F.}~\bibnamefont {van Wijland}},\ }\bibfield  {title} {\bibinfo {title} {Transverse forces and glassy liquids in infinite dimensions},\ }\href {https://doi.org/10.1103/PhysRevE.109.064133} {\bibfield  {journal} {\bibinfo  {journal} {Phys. Rev. E}\ }\textbf {\bibinfo {volume} {109}},\ \bibinfo {pages} {064133} (\bibinfo {year} {2024}{\natexlab{a}})}\BibitemShut {NoStop}%
\bibitem [{\citenamefont {Ghimenti}\ \emph {et~al.}(2024{\natexlab{b}})\citenamefont {Ghimenti}, \citenamefont {Berthier}, \citenamefont {Szamel},\ and\ \citenamefont {van Wijland}}]{Ghimenti2024b}%
  \BibitemOpen
  \bibfield  {author} {\bibinfo {author} {\bibfnamefont {F.}~\bibnamefont {Ghimenti}}, \bibinfo {author} {\bibfnamefont {L.}~\bibnamefont {Berthier}}, \bibinfo {author} {\bibfnamefont {G.}~\bibnamefont {Szamel}},\ and\ \bibinfo {author} {\bibfnamefont {F.}~\bibnamefont {van Wijland}},\ }\bibfield  {title} {\bibinfo {title} {Irreversible boltzmann samplers in dense liquids: Weak-coupling approximation and mode-coupling theory},\ }\href {https://doi.org/10.1103/PhysRevE.110.034604} {\bibfield  {journal} {\bibinfo  {journal} {Phys. Rev. E}\ }\textbf {\bibinfo {volume} {110}},\ \bibinfo {pages} {034604} (\bibinfo {year} {2024}{\natexlab{b}})}\BibitemShut {NoStop}%
\bibitem [{\citenamefont {Murugan}\ and\ \citenamefont {Vaikuntanathan}(2017)}]{murugan2017topologically}%
  \BibitemOpen
  \bibfield  {author} {\bibinfo {author} {\bibfnamefont {A.}~\bibnamefont {Murugan}}\ and\ \bibinfo {author} {\bibfnamefont {S.}~\bibnamefont {Vaikuntanathan}},\ }\bibfield  {title} {\bibinfo {title} {Topologically protected modes in non-equilibrium stochastic systems},\ }\href {https://doi.org/10.1038/ncomms13881} {\bibfield  {journal} {\bibinfo  {journal} {Nat. Commun.}\ }\textbf {\bibinfo {volume} {8}},\ \bibinfo {pages} {13881} (\bibinfo {year} {2017})}\BibitemShut {NoStop}%
\bibitem [{\citenamefont {Tang}\ \emph {et~al.}(2021)\citenamefont {Tang}, \citenamefont {Agudo-Canalejo},\ and\ \citenamefont {Golestanian}}]{Tang2021}%
  \BibitemOpen
  \bibfield  {author} {\bibinfo {author} {\bibfnamefont {E.}~\bibnamefont {Tang}}, \bibinfo {author} {\bibfnamefont {J.}~\bibnamefont {Agudo-Canalejo}},\ and\ \bibinfo {author} {\bibfnamefont {R.}~\bibnamefont {Golestanian}},\ }\bibfield  {title} {\bibinfo {title} {Topology protects chiral edge currents in stochastic systems},\ }\href {https://doi.org/10.1103/PhysRevX.11.031015} {\bibfield  {journal} {\bibinfo  {journal} {Phys. Rev. X}\ }\textbf {\bibinfo {volume} {11}},\ \bibinfo {pages} {031015} (\bibinfo {year} {2021})}\BibitemShut {NoStop}%
\bibitem [{\citenamefont {Ezawa}(2022)}]{Ezawa2022}%
  \BibitemOpen
  \bibfield  {author} {\bibinfo {author} {\bibfnamefont {M.}~\bibnamefont {Ezawa}},\ }\bibfield  {title} {\bibinfo {title} {Dynamical nonlinear higher-order non-hermitian skin effects and topological trap-skin phase},\ }\href {https://doi.org/10.1103/PhysRevB.105.125421} {\bibfield  {journal} {\bibinfo  {journal} {Phys. Rev. B}\ }\textbf {\bibinfo {volume} {105}},\ \bibinfo {pages} {125421} (\bibinfo {year} {2022})}\BibitemShut {NoStop}%
\bibitem [{\citenamefont {Nelson}\ and\ \citenamefont {Tang}(2024)}]{Nelson2024}%
  \BibitemOpen
  \bibfield  {author} {\bibinfo {author} {\bibfnamefont {A.}~\bibnamefont {Nelson}}\ and\ \bibinfo {author} {\bibfnamefont {E.}~\bibnamefont {Tang}},\ }\bibfield  {title} {\bibinfo {title} {Nonreciprocity is necessary for robust dimensional reduction and strong responses in stochastic topological systems},\ }\href {https://doi.org/10.1103/PhysRevB.110.155116} {\bibfield  {journal} {\bibinfo  {journal} {Phys. Rev. B}\ }\textbf {\bibinfo {volume} {110}},\ \bibinfo {pages} {155116} (\bibinfo {year} {2024})}\BibitemShut {NoStop}%
\bibitem [{\citenamefont {Zheng}\ and\ \citenamefont {Tang}(2024)}]{zheng2024topological}%
  \BibitemOpen
  \bibfield  {author} {\bibinfo {author} {\bibfnamefont {C.}~\bibnamefont {Zheng}}\ and\ \bibinfo {author} {\bibfnamefont {E.}~\bibnamefont {Tang}},\ }\bibfield  {title} {\bibinfo {title} {A topological mechanism for robust and efficient global oscillations in biological networks},\ }\href {https://doi.org/10.1038/s41467-024-50510-x} {\bibfield  {journal} {\bibinfo  {journal} {Nat. Commun.}\ }\textbf {\bibinfo {volume} {15}},\ \bibinfo {pages} {6453} (\bibinfo {year} {2024})}\BibitemShut {NoStop}%
\bibitem [{\citenamefont {Veenstra}\ \emph {et~al.}(2024)\citenamefont {Veenstra}, \citenamefont {Gamayun}, \citenamefont {Guo}, \citenamefont {Sarvi}, \citenamefont {Meinersen},\ and\ \citenamefont {Coulais}}]{veenstra2024non}%
  \BibitemOpen
  \bibfield  {author} {\bibinfo {author} {\bibfnamefont {J.}~\bibnamefont {Veenstra}}, \bibinfo {author} {\bibfnamefont {O.}~\bibnamefont {Gamayun}}, \bibinfo {author} {\bibfnamefont {X.}~\bibnamefont {Guo}}, \bibinfo {author} {\bibfnamefont {A.}~\bibnamefont {Sarvi}}, \bibinfo {author} {\bibfnamefont {C.~V.}\ \bibnamefont {Meinersen}},\ and\ \bibinfo {author} {\bibfnamefont {C.}~\bibnamefont {Coulais}},\ }\bibfield  {title} {\bibinfo {title} {Non-reciprocal topological solitons in active metamaterials},\ }\href {https://doi.org/10.1038/s41586-024-07097-6} {\bibfield  {journal} {\bibinfo  {journal} {Nature}\ }\textbf {\bibinfo {volume} {627}},\ \bibinfo {pages} {528} (\bibinfo {year} {2024})}\BibitemShut {NoStop}%
\bibitem [{\citenamefont {Belyansky}\ \emph {et~al.}(2025)\citenamefont {Belyansky}, \citenamefont {Weis}, \citenamefont {Hanai}, \citenamefont {Littlewood},\ and\ \citenamefont {Clerk}}]{Belyansky2025}%
  \BibitemOpen
  \bibfield  {author} {\bibinfo {author} {\bibfnamefont {R.}~\bibnamefont {Belyansky}}, \bibinfo {author} {\bibfnamefont {C.}~\bibnamefont {Weis}}, \bibinfo {author} {\bibfnamefont {R.}~\bibnamefont {Hanai}}, \bibinfo {author} {\bibfnamefont {P.~B.}\ \bibnamefont {Littlewood}},\ and\ \bibinfo {author} {\bibfnamefont {A.~A.}\ \bibnamefont {Clerk}},\ }\bibfield  {title} {\bibinfo {title} {Phase transitions in nonreciprocal driven-dissipative condensates},\ }\href {https://doi.org/10.1103/gphr-d1bc} {\bibfield  {journal} {\bibinfo  {journal} {Phys. Rev. Lett.}\ }\textbf {\bibinfo {volume} {135}},\ \bibinfo {pages} {123401} (\bibinfo {year} {2025})}\BibitemShut {NoStop}%
\bibitem [{\citenamefont {Frohoff-Hülsmann}\ \emph {et~al.}(2023)\citenamefont {Frohoff-Hülsmann}, \citenamefont {Holl}, \citenamefont {Knobloch}, \citenamefont {Gurevich},\ and\ \citenamefont {Thiele}}]{Frohoff2023}%
  \BibitemOpen
  \bibfield  {author} {\bibinfo {author} {\bibfnamefont {T.}~\bibnamefont {Frohoff-Hülsmann}}, \bibinfo {author} {\bibfnamefont {M.~P.}\ \bibnamefont {Holl}}, \bibinfo {author} {\bibfnamefont {E.}~\bibnamefont {Knobloch}}, \bibinfo {author} {\bibfnamefont {S.~V.}\ \bibnamefont {Gurevich}},\ and\ \bibinfo {author} {\bibfnamefont {U.}~\bibnamefont {Thiele}},\ }\bibfield  {title} {\bibinfo {title} {Stationary broken parity states in active matter models},\ }\href {https://doi.org/10.1103/PhysRevE.107.064210} {\bibfield  {journal} {\bibinfo  {journal} {Phys. Rev. E}\ }\textbf {\bibinfo {volume} {107}},\ \bibinfo {pages} {064210} (\bibinfo {year} {2023})}\BibitemShut {NoStop}%
\bibitem [{\citenamefont {Greve}\ \emph {et~al.}(2025)\citenamefont {Greve}, \citenamefont {Lovato}, \citenamefont {Frohoff-Hülsmann},\ and\ \citenamefont {Thiele}}]{Greve2025}%
  \BibitemOpen
  \bibfield  {author} {\bibinfo {author} {\bibfnamefont {D.}~\bibnamefont {Greve}}, \bibinfo {author} {\bibfnamefont {G.}~\bibnamefont {Lovato}}, \bibinfo {author} {\bibfnamefont {T.}~\bibnamefont {Frohoff-Hülsmann}},\ and\ \bibinfo {author} {\bibfnamefont {U.}~\bibnamefont {Thiele}},\ }\bibfield  {title} {\bibinfo {title} {Coexistence of uniform and oscillatory states resulting from nonreciprocity and conservation laws},\ }\href {https://doi.org/10.1103/PhysRevLett.134.018303} {\bibfield  {journal} {\bibinfo  {journal} {Phys. Rev. Lett.}\ }\textbf {\bibinfo {volume} {134}},\ \bibinfo {pages} {018303} (\bibinfo {year} {2025})}\BibitemShut {NoStop}%
\bibitem [{\citenamefont {Hohenberg}\ and\ \citenamefont {Halperin}(1977)}]{Hohenberg1977}%
  \BibitemOpen
  \bibfield  {author} {\bibinfo {author} {\bibfnamefont {P.~C.}\ \bibnamefont {Hohenberg}}\ and\ \bibinfo {author} {\bibfnamefont {B.~I.}\ \bibnamefont {Halperin}},\ }\bibfield  {title} {\bibinfo {title} {Theory of dynamic critical phenomena},\ }\href {https://doi.org/10.1103/RevModPhys.49.435} {\bibfield  {journal} {\bibinfo  {journal} {Rev. Mod. Phys.}\ }\textbf {\bibinfo {volume} {49}},\ \bibinfo {pages} {435} (\bibinfo {year} {1977})}\BibitemShut {NoStop}%
\bibitem [{\citenamefont {Frohoff-Hülsmann}\ and\ \citenamefont {Thiele}(2023)}]{Frohoff20232}%
  \BibitemOpen
  \bibfield  {author} {\bibinfo {author} {\bibfnamefont {T.}~\bibnamefont {Frohoff-Hülsmann}}\ and\ \bibinfo {author} {\bibfnamefont {U.}~\bibnamefont {Thiele}},\ }\bibfield  {title} {\bibinfo {title} {Nonreciprocal cahn-hilliard model emerges as a universal amplitude equation},\ }\href {https://doi.org/10.1103/PhysRevLett.131.107201} {\bibfield  {journal} {\bibinfo  {journal} {Phys. Rev. Lett.}\ }\textbf {\bibinfo {volume} {131}},\ \bibinfo {pages} {107201} (\bibinfo {year} {2023})}\BibitemShut {NoStop}%
\bibitem [{\citenamefont {Greve}\ and\ \citenamefont {Thiele}(2024)}]{Greve2024}%
  \BibitemOpen
  \bibfield  {author} {\bibinfo {author} {\bibfnamefont {D.}~\bibnamefont {Greve}}\ and\ \bibinfo {author} {\bibfnamefont {U.}~\bibnamefont {Thiele}},\ }\bibfield  {title} {\bibinfo {title} {An amplitude equation for the conserved-hopf bifurcation—derivation, analysis, and assessment},\ }\href {https://doi.org/10.1063/5.0222013} {\bibfield  {journal} {\bibinfo  {journal} {Chaos}\ }\textbf {\bibinfo {volume} {34}} (\bibinfo {year} {2024})}\BibitemShut {NoStop}%
\bibitem [{Note6()}]{Note6}%
  \BibitemOpen
  \bibinfo {note} {This contrasts \protect \textit {spurious} gradient dynamics (observed in some models with nonreciprocal $\protect \mathbf {H}$~\cite {Frohoff2023, Greve2025}) where $\protect \mathbf {L}$ and $\protect \mathbf {H}$ can be redefined such that both are symmetric but indefinite so that $\protect \mathbf {L}^{-1}$ gives a semi-Riemannian metric underlying a spurious gradient flow. This can occur in non-CB steady states if $\protect \bm {\protect \mathcal {V}}$ is symmetric but indefinite.}\BibitemShut {Stop}%
\bibitem [{\citenamefont {Sylvester}(1852)}]{Sylvester1852}%
  \BibitemOpen
  \bibfield  {author} {\bibinfo {author} {\bibfnamefont {J.~J.}\ \bibnamefont {Sylvester}},\ }\bibfield  {title} {\bibinfo {title} {Xix. a demonstration of the theorem that every homogeneous quadratic polynomial is reducible by real orthogonal substitutions to the form of a sum of positive and negative squares},\ }\href {https://doi.org/10.1080/14786445208647087} {\bibfield  {journal} {\bibinfo  {journal} {London Edinburgh Philos. Mag. J. Sci.}\ }\textbf {\bibinfo {volume} {4}},\ \bibinfo {pages} {138} (\bibinfo {year} {1852})}\BibitemShut {NoStop}%
\bibitem [{\citenamefont {Cahn}\ and\ \citenamefont {Hilliard}(1958)}]{Cahn1958}%
  \BibitemOpen
  \bibfield  {author} {\bibinfo {author} {\bibfnamefont {J.~W.}\ \bibnamefont {Cahn}}\ and\ \bibinfo {author} {\bibfnamefont {J.~E.}\ \bibnamefont {Hilliard}},\ }\bibfield  {title} {\bibinfo {title} {Free energy of a nonuniform system. i. interfacial free energy},\ }\href {https://doi.org/10.1063/1.1744102} {\bibfield  {journal} {\bibinfo  {journal} {J. Chem. Phys.}\ }\textbf {\bibinfo {volume} {28}},\ \bibinfo {pages} {258} (\bibinfo {year} {1958})}\BibitemShut {NoStop}%
\bibitem [{\citenamefont {Turing}(1952)}]{Turing1952}%
  \BibitemOpen
  \bibfield  {author} {\bibinfo {author} {\bibfnamefont {A.~M.}\ \bibnamefont {Turing}},\ }\bibfield  {title} {\bibinfo {title} {The chemical basis of morphogenesis},\ }\href {https://doi.org/10.1098/rstb.1952.0012} {\bibfield  {journal} {\bibinfo  {journal} {Phil. Trans. R. Soc. Lond. B}\ }\textbf {\bibinfo {volume} {237}},\ \bibinfo {pages} {37} (\bibinfo {year} {1952})}\BibitemShut {NoStop}%
\bibitem [{\citenamefont {Varga}(2011)}]{Varga2011}%
  \BibitemOpen
  \bibfield  {author} {\bibinfo {author} {\bibfnamefont {R.~S.}\ \bibnamefont {Varga}},\ }\href {https://link.springer.com/book/10.1007/978-3-642-17798-9} {\emph {\bibinfo {title} {Ger{\v{s}}gorin and his circles}}},\ Vol.~\bibinfo {volume} {36}\ (\bibinfo  {publisher} {Springer Science \& Business Media},\ \bibinfo {year} {2011})\BibitemShut {NoStop}%
\bibitem [{\citenamefont {Chung}(2005)}]{Chung2005}%
  \BibitemOpen
  \bibfield  {author} {\bibinfo {author} {\bibfnamefont {F.}~\bibnamefont {Chung}},\ }\bibfield  {title} {\bibinfo {title} {Laplacians and the cheeger inequality for directed graphs},\ }\href {https://doi.org/10.1007/s00026-005-0237-z} {\bibfield  {journal} {\bibinfo  {journal} {Ann. Comb.}\ }\textbf {\bibinfo {volume} {9}},\ \bibinfo {pages} {1} (\bibinfo {year} {2005})}\BibitemShut {NoStop}%
\end{thebibliography}
\end{document}


\maketitle

\pagenumbering{gobble}

\setcounter{secnumdepth}{2}
\setcounter{tocdepth}{2}
{
	\hypersetup{
		linkcolor=Black,
		citecolor=Black
	}
	\vspace{-10pt}
	\tableofcontents
}

\noindent\rule{5.0cm}{0.4pt}
	{\footnotesize
	$\\$
	{
	    {$^\dag \,$}aomar@berkeley.edu}
	}

\newpage 

\pagenumbering{arabic}

\section{When does chemostatting correspond to a semi-grand canonical ensemble?}

Consider a system described by Eq.~(2) in the main text with no reactions ($\mathbf{j} \cdot \boldsymbol{\nu}=\mathbf{0}$) and generation terms arising from a chemostat with $\mathbf{S} \cdot (\boldsymbol{\mu} - \boldsymbol{\mu}^{\rm chemo})$ where $\mathbf{S}$ is negative semi-definite and $\boldsymbol{\mu}^{\rm chemo}$ is a vector of chemical potentials of the chemostatted species in the external reservoir:
\begin{equation}
    \label{eq:semigranddyn}
    \partial_t \boldsymbol{\rho} = \overline{\mathbf{L}} \cdot \sum_{\alpha=1}^d \partial^2_\alpha \boldsymbol{\mu} + \mathbf{S} \cdot (\boldsymbol{\mu} - \boldsymbol{\mu}^{\rm chemo}).
\end{equation}
This system will relax (instantaneously in the limit $\vert \vert \mathbf{S} \vert \vert \rightarrow \infty$) to a detailed-balanced (DB) steady state at chemical potentials $\boldsymbol{\mu} = \boldsymbol{\mu}^{\rm chemo}$, corresponding to a grand or semi-grand canonical ensemble.
More generally, in the presence of chemical reactions, when the dynamics we consider in Eq.~(2) in the main text admit a DB steady state, they describe the relaxation of a system to a semi-grand canonical ensemble that is open to the chemostatted species.

For some reaction networks, certain species can be chemostatted such that a DB steady state does not exist.
When this is the case, chemostatting instead drives the system to nonequilibrium steady states.
This raises the question: given a stoichiometric matrix of chemostatted species $\boldsymbol{\nu}^{\rm chemo}$, can the chemostat chemical potentials be chosen such that a DB steady state does not exist?
We first note that closed systems described by Eq.~(2) in the main text, as well as open systems where $\boldsymbol{\mu}^{\rm chemo}=\mathbf{0}$, necessarily have a DB steady state.
Using the rank-nullity theorem, it can be shown that changing a dynamically evolving species to an autonomously chemostatted one necessarily breaks a conservation law (removes a right null vector of $\boldsymbol{\nu}$) or leads to an emergent cycle (introduces a left null vector of $\boldsymbol{\nu}$)~\cite{Polettini2014}.
A DB steady state persists when chemostatting does not introduce any emergent cycles.
In the language of Ref.~\cite{Rao2016}, this is when the steady-state affinity of each emergent cycle $\epsilon$, $\mathcal{A}^{\rm SS}_\epsilon \equiv k_B T \sum_{r}^{n_r} c_{\epsilon r} \mathcal{A}^{\rm SS}_r$ where $c_{\epsilon r}$ is the $r$ component of the $\epsilon$th left eigenvector of $\boldsymbol{\nu}$ and $\mathcal{A}^{\rm SS}_r \equiv -\sum_i^{n_e}\nu_{ri} \mu^{\rm SS}_i - \sum_i^{n_c} \nu_{ri}^{\rm chemo} \mu_i^{\rm chemo}$ is the steady-state affinity of reaction $r$, vanishes.
We now frame these conditions using the notation used in our manuscript.

The conditions for a DB steady state, $\mathbf{j}^{\rm SS}=\mathbf{0}$, are equivalent to vanishing affinity for each reaction ($\mathcal{A}_r = 0~\forall \ r$):
\begin{equation}
    \label{eq:detbal}
    \boldsymbol{\nu} \cdot \boldsymbol{\mu} + \boldsymbol{\nu}^{\rm chemo} \cdot \boldsymbol{\mu}^{\rm chemo} = \mathbf{0}.
\end{equation}
If $\boldsymbol{\nu}$ has at least one right null vector (corresponding to a conservation law), these conditions can be met if $\boldsymbol{\mu}^{\rm chemo}$ is a right null vector of $\boldsymbol{\nu}^{\rm chemo}$.
This corresponds to a conservation law of a closed system described by the stoichiometry of the chemostatted species.
More generally, the conditions to retain DB steady states can be found by contracting the matrix whose rows are left null vectors of $\boldsymbol{\nu}$, $\mathbf{N}$, with Eq.~\eqref{eq:detbal}:
\begin{equation}
    \label{eq:detbal2}
    \mathbf{N} \cdot \boldsymbol{\nu}^{\rm chemo} \cdot \boldsymbol{\mu}^{\rm chemo} = \mathbf{0},
\end{equation}
where we have made use of $\mathbf{N} \cdot \boldsymbol{\nu} = \mathbf{0}$.
Therefore, to retain a DB steady state, $\boldsymbol{\mu}^{\rm chemo}$ must be a right null vector of $\mathbf{N} \cdot \boldsymbol{\nu}^{\rm chemo}$.
As each element of both $\mathbf{N}$ and $\boldsymbol{\nu}^{\rm chemo}$ are integers, the mandated form of $\boldsymbol{\mu}^{\rm chemo}$ corresponds to a conservation law of a closed system described by the stoichiometry corresponding to the matrix $\mathbf{N} \cdot \boldsymbol{\nu}^{\rm chemo}$.

\section{Lyapunov functional for dynamics near open detailed-balanced systems\label{sec:openlya}}
We now describe the approach used in Ref.~\cite{Avanzini2024} to derive the Lyapunov functional of open reaction networks where $\boldsymbol{\nu}$ has no left null space (i.e.,~there are no reaction cycles) or, in the language of Ref.~\cite{Avanzini2024}, ``pseudo'' DB systems.
We begin by defining $n_\epsilon$ ``lumped'' reactions, each denoted $\epsilon$ and having an associated set of reaction $\mathcal{R}_\epsilon$ that holds reactions with the same stoichiometry, i.e.,~$\nu_{\epsilon i}=\nu_{r i} = \nu_{r' i} \ \forall \ r, r' \in \mathcal{R}_\epsilon$.
For example, the two reactions $A \rightleftharpoons B$ and $A \rightleftharpoons B + C$ fall under a single lumped reaction when $C$ is chemostatted.

The approach in Ref.~\cite{Avanzini2024} is to seek a mapping from an open system to a closed (non-chemostatted) system with equivalent reaction rates whose stoichiometry matrix has components $\nu_{\epsilon i}$ and whose chemical potentials, $\overline{\mu}_i = \mu_i + c_i$, are the chemical potentials of the open system, $\mu_i$, shifted by some constant, $c_i$.
In particular:
\begin{equation}
    k_\epsilon \exp\left(\beta \sum_i^{n_e} \nu_{\epsilon i}^\pm \overline{\mu}_i\right) = \sum_{r \in \mathcal{R}_\epsilon} k_r \exp \left(\beta \sum_i^{n_e} \nu_{r i}^\pm \mu_i + \beta \sum_i^{n_c} \nu_{r i}^{\pm, \rm chemo} \mu_i^{\rm chemo} \right).
\end{equation}
It follows that:
\begin{subequations}
    \begin{align}
        \sum_i^{n_e} \nu_{\epsilon i} c_i =& \log \left( \frac{\sum_{r \in \mathcal{R}_\epsilon} k_r \exp\left(\sum_i^{n_c} \nu_{r i}^{-, \rm chemo} \mu_i^{\rm chemo}\right)}{\sum_{r \in \mathcal{R}_\epsilon} k_r \exp\left(\sum_i^{n_c} \nu_{r i}^{+, \rm chemo} \mu_i^{\rm chemo}\right)} \right), \label{eq:cieqs} \\
        k_\epsilon^2  =& \sum_{r, r' \in \mathcal{R}_\epsilon} k_r k_{r'} \exp \left(\sum_i^{n_c} \left(\nu_{r i}^{-, \rm chemo} + \nu_{r' i}^{+, \rm chemo} \right) \mu_i^{\rm chemo} - \sum_i^{n_e}\left( \nu_{\epsilon i}^+ + \nu_{\epsilon i}^- \right) c_i\right). \label{eq:kepseq}
    \end{align}
\end{subequations}
Noting that $n_e>n_\epsilon$ necessarily due to the lack of reaction cycles, one can always define two subsets of species, $\mathcal{I}$ and $\mathcal{D}$ where $c_i = c_i^{\rm sub} \ \forall \ i \in \mathcal{I}$ and $c_i = 0 \ \forall \ i \in \mathcal{D}$, such that the matrix with components $\nu_{\epsilon i}^{\rm sub}=\nu_{\epsilon i} \ \forall \ i \in \mathcal{I}$ is square and invertible.
Eq.~\eqref{eq:cieqs} can then be solved with:
\begin{equation}
    \label{eq:cisubsolution}
    c_i^{\rm sub} = \sum_\epsilon^{n_\epsilon}\left(\nu_{i \epsilon}^{\rm sub}\right)^{-1}\log \left( \frac{\sum_{r \in \mathcal{R}_\epsilon} k_r \exp\left(\sum_i^{n_c} \nu_{r i}^{-, \rm chemo} \mu_i^{\rm chemo}\right)}{\sum_{r \in \mathcal{R}_\epsilon} k_r \exp\left(\sum_i^{n_c} \nu_{r i}^{+, \rm chemo} \mu_i^{\rm chemo}\right) } \right),
\end{equation}
and the resulting solution can be substituted into Eq.~\eqref{eq:kepseq} to solve for $k_\epsilon$.
The free energy describing this effectively closed system:
\begin{equation}
    \label{eq:Fbardef}
    \overline{F} \equiv F + \mathbf{c} \cdot \int d \mathbf{x} \boldsymbol{\rho}(\mathbf{x}),
\end{equation}
is hence a Lyapunov functional and governs the steady states of the system.
Equivalently, steady states are governed by the semi-grand canonical potential:
\begin{equation}
    \label{eq:omegasemi}
    \Omega^{\rm semi} \equiv F - \boldsymbol{\mu}^{\rm chemo} \cdot \mathbf{N}^{\rm chemo}.
\end{equation}
While both Eq.~\eqref{eq:Fbardef} and Eq.~\eqref{eq:omegasemi} can be used to identify the steady-state equilibrium concentrations the species adopt, we emphasize that the semi-grand potential in Eq.~\eqref{eq:omegasemi} is \textit{not} a Lyapunov functional for the dynamics, as the supply/removal of autonomously chemostatted species generally can increase $\Omega^{\rm semi}$ over a finite time interval.

\section{Coercive dynamics near detailed and complex-balance steady states}

We now briefly demonstrate that the linearized dynamics near stable DB and complex-balanced (CB) steady states are always coercive with respect to a Riemannian metric.
To see this, we first note that $\boldsymbol{\mathcal{V}}$ can be written as $\boldsymbol{\Gamma}^{\top} \cdot \boldsymbol{\mathcal{V}}^{\rm complex} \cdot \boldsymbol{\Gamma}$ where the components of $\boldsymbol{\mathcal{V}}^{\rm complex}$ are defined in Eq.~(13) in the main text.
In DB and CB steady states, $\boldsymbol{\mathcal{V}}^{\rm complex}$ is respectively the Laplacian of an undirected and directed graph.
Consequently, the kernel of the symmetric part of $\boldsymbol{\mathcal{V}}$, $\boldsymbol{\mathcal{V}}^{\rm S}$, is a subset of the kernel of $\boldsymbol{\mathcal{V}}$.
Therefore, any null vector $\mathbf{u}$ satisfying $\boldsymbol{\mathcal{V}}^{\rm S} \cdot \mathbf{u} = \mathbf{0}$ necessarily satisfies $\boldsymbol{\mathcal{V}}^{\rm A} \cdot \mathbf{u} = \mathbf{0}$, where $\boldsymbol{\mathcal{V}}^{\rm A}$ is the antisymmetric part of $\boldsymbol{\mathcal{V}}$.

Taking the quadratic form of $\mathbf{A} = -\mathbf{L} \cdot \mathbf{H}$ using $\mathbf{H}^{-1}$ as a metric, assuming $\mathbf{H}$ is symmetric positive-definite (PD) as the steady state must be stable, we have:
\begin{equation}
    \left \langle \mathbf{v} \cdot \mathbf{A}, \mathbf{v} \right \rangle_{\mathbf{H}^{-1}} = -\mathbf{v} \cdot \mathbf{L}^{\rm S} \cdot \mathbf{v} \leq 0,
\end{equation}
where $\mathbf{L}^{\rm S}$ is the symmetric part of $\mathbf{L}$.
When $q>0$, the dynamics are trivially coercive with respect to this metric.
When $q=0$, the dynamics are again coercive with respect to this metric as any null vector of $\mathbf{L}^{\rm S}$ is also in the kernel of the antisymmetric part of $\mathbf{L}$.
It is therefore clear that if one prepares a thermodynamically stable DB or CB steady state, small fluctuations of all wavelengths decay exponentially quickly.

\section{Lyapunov functionals for linear force dynamics
\label{sec:linearforces}}

The full nonlinear dynamics presented in Eq.~(2) in the main text are anticipated to hold when the system remains in a state of local equilibrium.
We now consider the dynamics near a spatially uniform steady-state solution that are linear in \textit{driving forces}, i.e.,~chemical potential differences.
The spatially constant species densities are defined as $\boldsymbol{\rho}^{\rm SS}$ and the chemical potentials follow as $\boldsymbol{\mu}^{\rm SS} = \boldsymbol{\mu}(\boldsymbol{\rho}^{\rm SS})$. 
The departure of the concentrations and chemical potentials from their steady-state values are respectively defined as $\delta \boldsymbol{\rho} \equiv \boldsymbol{\rho} - \boldsymbol{\rho}^{\rm SS}$ and $\delta \boldsymbol{\mu} \equiv \boldsymbol{\mu} - \boldsymbol{\mu}^{\rm SS}$. 
We can linearize the species dynamics [Eq.~(2) in the main text] with respect to $\delta \boldsymbol{\mu}$:
\begin{equation}
\label{eq:rhodynlinmu}
    \partial_t \delta \boldsymbol{\rho} \approx \left( \boldsymbol{\mathcal{V}} + \sum_{\alpha=1}^d \partial_\alpha \overline{\mathbf{L}} \partial_\alpha  \right) \cdot \delta \boldsymbol{\mu},
\end{equation}
where the components of the $n_e \times n_e$ matrix $\boldsymbol{\mathcal{V}}$ are defined in Eq.~(3b) in the main text.

We now define the change in $\overline{F}$ (see Section~\ref{sec:openlya} for the definition of $\overline{F}$) due to $\delta \boldsymbol{\rho}$:
\begin{subequations}
    \begin{align}
        \delta \overline{F} \equiv& \overline{F}[\boldsymbol{\rho}(\mathbf{x})] - \overline{F}[\boldsymbol{\rho}^{\rm SS}] = \delta F + \mathbf{c} \cdot \int d \mathbf{x} \delta \boldsymbol{\rho}(\mathbf{x}), \\
        \delta F \equiv& F[\boldsymbol{\rho}(\mathbf{x})] - F[\boldsymbol{\rho}^{\rm SS}], \\
        \boldsymbol{\mu} =& \frac{\delta(\delta F)}{\delta (\delta \boldsymbol{\rho})} = \boldsymbol{\mu}^{\rm SS} + \delta \boldsymbol{\mu}.
    \end{align}
\end{subequations}
In DB steady states, we find:
\begin{subequations}
    \begin{align}
         \partial_t \delta \overline{F} =& \partial_t \delta \overline{F}^{\rm rxn} + \partial_t \delta \overline{F}^{\rm flux} \leq 0, \\
         \partial_t \delta \overline{F}^{\rm rxn} \approx& \int d \mathbf{x} \left( \boldsymbol{\mu}^{\rm SS} + \delta \boldsymbol{\mu} + \mathbf{c} \right) \cdot \boldsymbol{\mathcal{V}} \cdot \delta \boldsymbol{\mu} = \int d \mathbf{x} \delta \boldsymbol{\mu} \cdot \boldsymbol{\mathcal{V}} \cdot \delta \boldsymbol{\mu} \leq 0, \label{eq:omegasemirxn}\\ 
         \partial_t \delta \overline{F}^{\rm flux} =& \int d \mathbf{x} \left( \boldsymbol{\mu}^{\rm SS} + \delta \boldsymbol{\mu} + \mathbf{c} \right) \cdot \sum_{\alpha=1}^d \partial_\alpha \mathbf{L} \cdot \partial_\alpha \delta \boldsymbol{\mu} = -\int d \mathbf{x} \sum_{\alpha=1}^d \partial_\alpha \delta \boldsymbol{\mu} \cdot \overline{\mathbf{L}} \cdot \partial_\alpha \delta \boldsymbol{\mu} \leq 0, \label{eq:omegasemiflux}
    \end{align}
\end{subequations}
where we have used $( \boldsymbol{\mu}^{\rm SS} + \mathbf{c}) \cdot \boldsymbol{\mathcal{V}} = \mathbf{0}$ (see Section~\ref{sec:openlya}), the fact that $\boldsymbol{\mathcal{V}}$ is positive semi-definite (PSD) in DB steady states, and integrated by parts and used zero-flux boundary conditions in Eq.~\eqref{eq:omegasemiflux} (which applies to CB and non-CB steady states as well).
It is clear that $\delta \overline{F}$ is a Lyapunov functional for these linear (in $\delta \boldsymbol{\mu}$) dynamics.
Near CB steady states, we have:
\begin{subequations}
    \begin{align}
         \partial_t \delta \overline{\boldsymbol{\Omega}}=& \partial_t \delta \overline{\boldsymbol{\Omega}}^{\rm rxn} + \partial_t \delta \overline{\boldsymbol{\Omega}}^{\rm flux} \leq 0, \\
         \partial_t \delta \overline{\boldsymbol{\Omega}}^{\rm rxn} \approx& \int d \mathbf{x} \left( \boldsymbol{\mu}^{\rm SS} + \delta \boldsymbol{\mu} - \boldsymbol{\mu}^{\rm SS} \right) \cdot \boldsymbol{\mathcal{V}} \cdot \delta \boldsymbol{\mu} = \int d \mathbf{x} \delta \boldsymbol{\mu} \cdot \boldsymbol{\mathcal{V}} \cdot \delta \boldsymbol{\mu} \leq 0, \label{eq:omegarxn} \\ 
         \overline{\boldsymbol{\Omega}}^{\rm flux} =& \overline{F}^{\rm flux} \leq 0, \label{eq:omegaflux}
    \end{align}
\end{subequations}
where we have used the fact that the symmetric part of $\boldsymbol{\mathcal{V}}$ (which is what controls its quadratic form) is PSD in Eq.~\eqref{eq:omegarxn}.
As was the case for $\delta \overline{F}$ near DB steady states, $\delta \boldsymbol{\Omega}$ is clearly a Lyapunov functional near CB steady states.
When the symmetric part of $\boldsymbol{\mathcal{V}}$ is indefinite, as is the case in non-CB steady states, neither $\Omega$ nor $\overline{F}$ is a Lyapunov functional at the linear (or nonlinear) level.

\section{Non-Hermitian skin effects in chemical reaction chains} 

Consider the cyclic reaction networks studied in the main text with the final reaction removed, i.e.,~the $i$th (of $n_r = n_c = n_e - 1$ in total) takes the form ${Z_i \rightleftharpoons Z_{i + 1} + Z_{i}^{\rm chemo}}$.
For this network, $\boldsymbol{\nu}$ has no left null vectors and thus any steady state is necessarily DB.
As a result, $\mathbf{A}$ can no longer have complex eigenvalues.
Moreover, as discussed in Section~\ref{sec:openlya}, this system has a Lyapunov functional that corresponds to the free energy of a closed system with chemical potentials shifted by constants related to the chemostats.
By defining $k_r^\pm \equiv k_r \exp(\sum_i^{n_c} \nu_{r i}^{\pm, \rm chemo} \mu_i^{\rm chemo})$ (therefore $k_r^+ = k_r~\forall \ r$), we can express the dynamics of the concentration of the $1<i<n_e$th species (ignoring diffusion and gradient terms in the free energy) as:
\begin{subequations}
    \label{eq:rhoidyngamma}
    \begin{equation}
        \partial_t \rho_i = k_{i-1}^+ \exp\left(\beta \mu_{i-1}\right) + k_{i}^- \exp\left(\beta \mu_{i+1}\right) - \left( k_{i-1}^- + k_i^+ \right) \exp\left(\beta \mu_{i}\right),
    \end{equation}
    and the dynamics of the first and $n_e$th species as:
    \begin{align}
        \partial_t \rho_1 =& k_{1}^- \exp\left(\beta \mu_{2}\right) - k_1^+ \exp\left(\beta \mu_{1}\right), \\
        \partial_t \rho_{n_e} =& k_{n_e-1}^+ \exp\left(\beta \mu_{n_e-1}\right) - k_{n_e-1}^-\exp\left(\beta \mu_{n_e}\right).
    \end{align}
\end{subequations}

For mass action kinetics (i.e.,~for ideal solutions) where ${\mu_i=\log(\rho_i/\rho_i^{\rm ref})}$ for some reference density $\rho_i^{\rm ref}$, Eq.~\eqref{eq:rhoidyngamma} recovers a Hatano-Nelson model~\cite{Hatano1996, Hatano1997, Zhang2022} with open boundary conditions when $k_r^\pm = k^\pm \ \forall \ r$ (this implies $k_r = k \ \forall \ r$ and $\mu_i^{\rm chemo} = \mu^{\rm chemo}~\forall \ i$ as $k^+=k_r~\forall \ r$) and $\rho_i^{\rm ref}=\rho^{\rm ref} \ \forall \ i$.
To see this, we first note that the components of $\mathbf{H}$ are $H_{ij} = k_B T \delta_{ij} / \rho_i^{\rm SS}$, while those of $\boldsymbol{\mathcal{V}}$ are:
\begin{subequations}
    \begin{equation}
        \mathcal{V}_{ij} = \beta \left[ \delta_{ij} \left( j_i^{\pm, \rm SS} + j_{i-1}^{\pm, \rm SS} \right) - \delta_{(i-1)j} j_{i-1}^{\pm, \rm SS} - \delta_{i(j-1)} j_i^{\pm, \rm SS}\right]
    \end{equation}
    for $1 < i < n_e$ and for $i=1$ and $i=n_e$ they are:
    \begin{align}
        \mathcal{V}_{1j} =& \beta \left( \delta_{1j}j_1^{\pm, \rm SS} - \delta_{1(j-1)} j_1^{\pm, \rm SS}\right), \\
        \mathcal{V}_{n_ej} =& \beta\left( \delta_{n_ej} j_{n_e-1}^{\pm, \rm SS} - \delta_{(n_e-1)j} j_{n_e-1}^{\pm, \rm SS} \right).
    \end{align}
\end{subequations}
Consequently, $\mathbf{A}$ has components:
\begin{subequations}
    \begin{equation}
         A_{ij} = \frac{1}{\rho^{\rm ref}} \left[ k^+ \delta_{j(i-1)} + k^- \delta_{j(i+1)} -  \left( k^+ + k^- \right) \delta_{i j} \right]
    \end{equation}
    for $1 < i < n_e$ and a first and $n_e$th row with components:
    \begin{align}
        A_{1j} =& \frac{1}{\rho^{\rm ref}} \left( k^- \delta_{j2} -  k^+ \delta_{j1} \right), \\
        A_{n_e j} =& \frac{1}{\rho^{\rm ref}} \left( k^+ \delta_{j(n_e-1)} -  k^- \delta_{j n_e} \right).
    \end{align}
\end{subequations}
This Jacobian is asymmetric and has right eigenvectors (for the $0<m<n_e-1$ nonzero modes):
\begin{equation}
    R_i^{(m)} = \left( \sqrt{\frac{k^+}{k^-}} \right)^{i-1} \left[ \sin \left( \frac{\pi m i}{n_e} \right) - \sqrt{\frac{k^-}{k^+}} \sin \left( \frac{\pi m (i-1)}{n_e} \right) \right],
\end{equation}
that exponentially localize on the $i=n_e$ boundary when $k^+ > k^-$ and the $i=1$ boundary when $k^+ < k^-$ -- the hallmark of the non-Hermitian skin effect.
Importantly, as $\boldsymbol{\mathcal{V}}$ is symmetric PSD and $\mathbf{H}$ is symmetric PD, $\mathbf{A}$ is similar to a symmetric (Hermitian) matrix and thus its spectrum is strictly real, as anticipated.
This can be seen through the transformation we discuss in the main text which is equivalent to an ``imaginary gauge transformation''~\cite{Hatano1996, Hatano1997}.

These chemical non-Hermitian skin effects can be better understood through the mapping to a closed system discussed in Section~\ref{sec:openlya}, where the constant chemical potential shifts take the form~\footnote{The inverse matrix of the stoichiometric matrix (ignoring the first column), $(\boldsymbol{\nu}^{\rm sub})^{-1}$, is a lower triangular matrix of ones -- see Eq.~\eqref{eq:cisubsolution} for the general form of $\mathbf{c}$.}:
\begin{equation}
    c_i = -(i-1)k_B T \log \left( \frac{ k^+}{ k^- } \right) = (i-1)k_B T \mu^{\rm chemo},
\end{equation}
where we have used $k^+/k^-=\exp(- \mu^{\rm chemo})$ in the second equality.
The effectively closed system is then described by the free energy $\overline{F}\equiv F + \mathbf{c} \cdot \int d \mathbf{x} \boldsymbol{\rho}(\mathbf{x})$ which is a Lyapunov functional for these dynamics.
This constant shift, $\mathbf{c}$, corresponds to a change in the reference chemical potential (or, equivalently, reference density $\rho_i^{\rm ref}$) of each of the species -- this can arise from a difference in the molecular partition function of each species.
As $c_i$ decreases linearly with $i$ if $k^+>k^-$ ($\mu^{\rm chemo}<0$) and increases linearly if $k^+<k^-$ ($\mu^{\rm chemo}>0$), it is clear that this chemical Hatano-Nelson model corresponds to a chemical chain where the species have progressively lower ($k^+>k^-$) or higher ($k^+<k^-$) molecular free energies -- the exponential weighting near the boundaries is a result of the Boltzmann distribution.
It remains an open question what other non-Hermitian topological effects~\cite{murugan2017topologically,Tang2021,Zhang2022,Ezawa2022, Okuma2023, Nelson2024,zheng2024topological,veenstra2024non,Belyansky2025} can be reproduced in chemical reaction networks that minimize a free energy.

\section{Free energy minimization in cyclic networks}

Consider the $n=3$  cyclic reaction network illustrated in Fig.~1 in the main text, where $k_r=1$ and $j_r^-=0$ for all $r$.
The stoichiometry matrix takes the form:
\begin{equation}
    \boldsymbol{\nu} = \begin{bmatrix}
        -1 & 1 & 0 \\ 0 & -1 & 1 \\ 1 & 0 & -1
    \end{bmatrix},
\end{equation}
which has a null vector $\boldsymbol{u} = \begin{bmatrix} 1 & 1 & 1 \end{bmatrix}^{\top}$ (both left and right).
This means the reaction sequence $j_1 \rightarrow j_2 \rightarrow j_3$ corresponds to a cycle and the total density $\rho_{\rm tot}=\rho_A+\rho_B+\rho_C$ is conserved.
Introducing the Lagrangian $\mathcal{L} \equiv f^{\rm bulk} - \mu^{\rm coexist} (\rho_A + \rho_B + \rho_C - \rho_{\rm tot})$ where $f^{\rm bulk}$ is the bulk free energy density and $\mu^{\rm coexist}$ is a Lagrange multiplier, the criteria for the free energy to be minimized subject to total density conservation follow as:
\begin{equation}
     \frac{\partial \mathcal{L}}{\partial \rho_i} = 0 \implies \mu^{\rm bulk}_i = \mu^{\rm coexist},
\end{equation}
where $\mu^{\rm bulk}_i \equiv \partial f^{\rm bulk} / \partial \rho_i$
for each species $i$ along with the constraint $\rho_A + \rho_B + \rho_C = \rho_{\rm tot}$.
We then see that $\mu^{\rm bulk}_A=\mu^{\rm bulk}_B=\mu^{\rm bulk}_C$ is both the thermodynamic criteria for free energy minimization and the dynamic criteria for complex-balancing in this system.
Notably, as $\boldsymbol{\mu}^{\rm SS} = \mathbf{0}$, the free energy is equivalent to the grand potential (up to a constant $\boldsymbol{\mu}^{\rm chemo} \cdot \mathbf{N}^{\rm chemo}$).
\bigskip

An alternative formulation of the thermodynamic criteria can be found by directly imposing $\rho_C = \rho_{\rm tot} - \rho_A - \rho_B$ for some fixed $\rho_{\rm tot}$.
For the expanded (i.e.,~near a steady state) free energy used to construct Fig.~2 in the main text, $\beta vf^{\rm bulk}=\sum_i^{n_e}(-\phi_i/2+\phi_i^4/4)$ where $\phi_i = v(\rho_i - \rho^{\rm ref}_i)$ is the shifted concentration of species $i$ ($v$ is a characteristic volume), this leads to:
\begin{equation}
    \beta vf^{\rm bulk} = -\frac{\phi_A^2}{2}+\frac{\phi_A^4}{4} -\frac{\phi_B^2}{2}+\frac{\phi_B^4}{4} -\frac{(\phi_{\rm tot} - \phi_B - \phi_A)^2}{2}+\frac{(\phi_{\rm tot} - \phi_A - \phi_B)^4}{4}.
\end{equation}
The free energy can now by minimized without introducing a Lagrange multiplier, leading to:
\begin{subequations}
    \label{eq:muzero}
    \begin{align}
        \mu_A^{\rm bulk} = -2\phi_A - \phi_B + \phi_A^3 + \phi_{\rm tot} - (\phi_{\rm tot} - \phi_A - \phi_B)^3 = 0, \\
        \mu_B^{\rm bulk} = -2\phi_B - \phi_A + \phi_B^3 + \phi_{\rm tot} - (\phi_{\rm tot} - \phi_A - \phi_B)^3 = 0,
    \end{align}
\end{subequations}
which must be supplemented by mandating the Hessian with elements $H_{ij} \equiv \partial^2 f^{\rm bulk} / \partial \rho_i \partial \rho_j$ is PD.

Solutions $\phi_A^{*}$ and $\phi_B^{*}$ that solve Eq.~\eqref{eq:muzero} can determined analytically.
The free energy at the minima prescribed by $\phi_A^{*}$ and $\phi_B^{*}$, $f^*$, takes the following $\phi_{\rm tot}$-dependent form:
\begin{equation}
    \label{eq:fstar}
    \beta vf^* = -\frac{1}{2} - \frac{\phi_{\rm tot} \sqrt{12 - 3\phi_{\rm tot}^2}}{9} + \frac{\phi_{\rm tot}^3 \sqrt{12 - 3\phi_{\rm tot}^2}}{36},
\end{equation}
when $0 < \phi_{\rm tot} < \sqrt{3}$ with $f^* \rightarrow -f^*$ when $\phi_{\rm tot} \rightarrow -\phi_{\rm tot}$.

\begin{figure}
    \centering
    \includegraphics[width=0.82\linewidth]{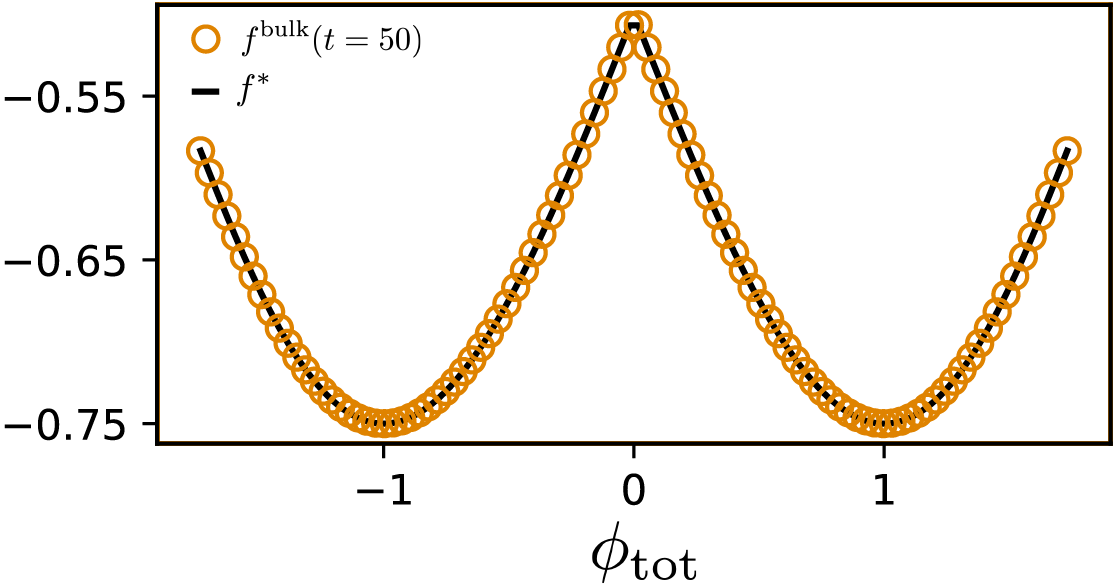}
    \caption{The free energy of a three-component homogeneous system prepared at $\phi_i=\phi_{\rm tot}/n$ (with noise), where $\beta vf^{\rm bulk} = \sum_i^{3}(-\phi^2_i/2+\phi_i^4/4)$, after evolving for a time $t=50$ (orange circles) compared to the minimum free energy, $f^*$, determined through Eq.~\eqref{eq:fstar} (black line). The free energy densities are presented in units of $(\beta v)^{-1}$.}
    \label{fig:n3f}
\end{figure}

In Fig.~\ref{fig:n3f}, we compare $f^*$ with the system free energy found by simulating a homogeneous system.
We allow the system to evolve from an unstable steady state at $\phi_i=\phi_{\rm tot}/3 \ \forall \ i$ until a stable steady state is reached, finding excellent agreement between the free energy at the stable steady state and Eq.~\eqref{eq:fstar}.
This makes clear that in cyclic reaction networks, such as those in Figs.~1 and 2 in the main text, the value of the total concentration determines the free energy minima the system will evolve to.

\section{Analogy to overdamped particle in a potential}
\begin{figure*}
    \centering
    \includegraphics[width=0.99\linewidth]{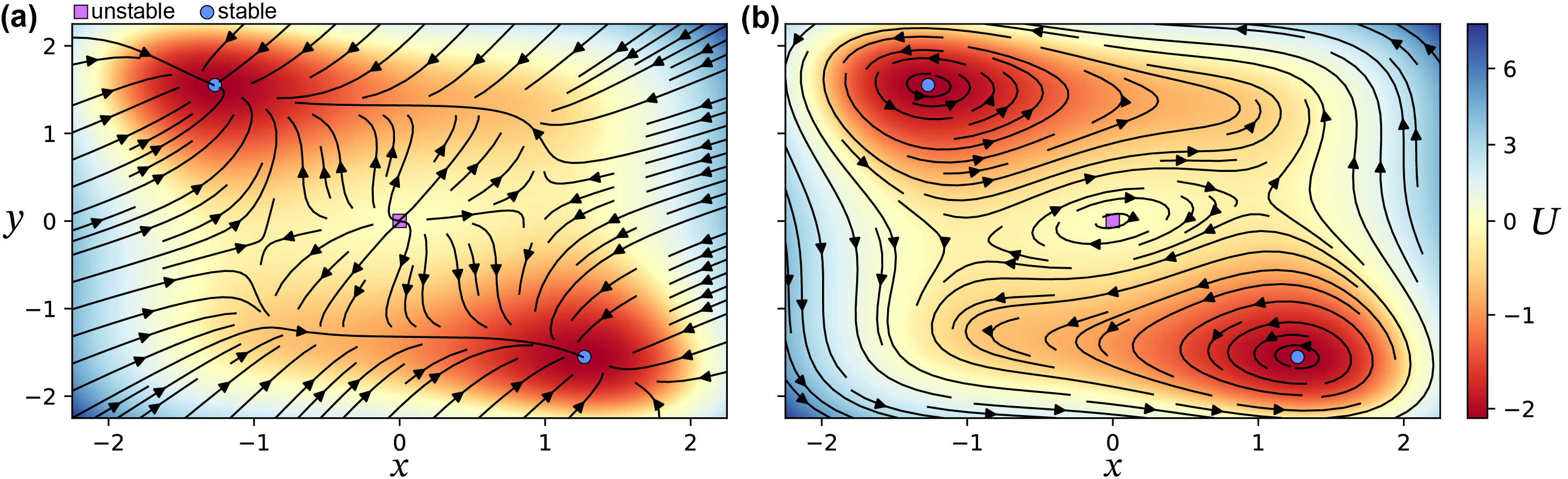}
    \caption{Phase planes of an overdamped particle in a potential $U = -x^2/2 + x^4/4 - y^2 + y^4/4 + xy/2$ where the drag $\boldsymbol{\zeta}$ is (a) even with the form $\boldsymbol{\zeta} = \begin{bmatrix} 1 & 1/2 \\ 1/2 & 1 \end{bmatrix}$; and (b) odd with the form $\boldsymbol{\zeta} = \begin{bmatrix} 1/10 & 1 \\ -1 & 1/10 \end{bmatrix}$. Stable and unstable steady states are marked with blue circles and purple squares, respectively.}
    \label{fig:oddhill}
\end{figure*}

Nonreciprocity in our chemical networks while retaining a Lyapunov functional shares similarities to single-particle dynamics in the presence of ``odd'' transport coefficients. 
Consider a particle in a two-dimensional potential that is bounded from below, $U(x, y)$, with position $\mathbf{r} \equiv \begin{bmatrix} x & y \end{bmatrix}^{\top}$ evolving according to the dynamics as:
\begin{equation}
    \label{eq:rdot}
    \partial_t \mathbf{r} = - \boldsymbol{\zeta} \cdot \boldsymbol{\nabla} U,
\end{equation}
where $\boldsymbol{\nabla} \equiv \begin{bmatrix} \partial / \partial x & \partial / \partial y \end{bmatrix}^{\top}$ and $\boldsymbol{\zeta}$ is a $2 \times 2$ matrix whose symmetric part, $\boldsymbol{\zeta}^{\rm S} \equiv (\boldsymbol{\zeta} + \boldsymbol{\zeta}^{\top})/2$, is positive-definite.
This ensures that $U$ is a Lyapunov functional for these dynamics:
\begin{equation}
    \partial_t U = - \boldsymbol{\nabla} U \cdot \boldsymbol{\zeta}^{\rm S} \cdot \boldsymbol{\nabla} U \leq 0,
\end{equation}
where $\partial_t U=0$ if and only if $\boldsymbol{\nabla} U=\mathbf{0}$, precluding limit cycles.
Importantly, this means that the drag may be nonreciprocal or \textit{odd} (as is the case for a tracer particle in a chiral active fluid~\cite{poggioli2023}, for example), $\boldsymbol{\zeta} \neq \boldsymbol{\zeta}^{\top}$, yet $U$ will remain a Lyapunov functional as long as $\boldsymbol{\zeta}^{\rm S}$ is positive-definite.
In Fig.~\ref{fig:oddhill} we simulate this system with (a) even and (b) odd drag, demonstrating that the steady states and their overall stability remain the same.
However, temporal oscillations occur around the unstable steady state in (b), corresponding to a Hopf instability where the Jacobian has eigenvalues $\approx 0.15 \pm 1.32 i$.
Furthermore, beyond the region of linear instability the particle in (b) oscillates around the potential while settling into a minima.

We now comment on the structure of Eqs.~\eqref{eq:rdot} and \eqref{eq:rhodynlinmu}.
Indeed, we recognize that the driving forces $\boldsymbol{\nabla}U$ and $\delta \boldsymbol{\mu}$ are both variational.
This means that the definiteness (and null space) of the transport matrices, $\boldsymbol{\zeta}$ and $\boldsymbol{\mathcal{V}}$, controls the existence of a Lyapunov functional.
In systems with odd drag, $\boldsymbol{\zeta}$ is asymmetric but its symmetric part is PD, causing $U$ to remain a Lyapunov functional despite the presence of spatial oscillations.
Analogously, near CB steady states, $\boldsymbol{\mathcal{V}}$ is asymmetric but its symmetric part is PSD.
This causes the appropriate free energy (the grand potential) to remain a Lyapunov functional despite temporal oscillations in the species concentrations.

\addcontentsline{toc}{section}{References}
\bibliographystyle{bibStyle}